\newcommand{\beq}{\begin{eqnarray}}
\newcommand{\eeq}{\end{eqnarray}}

\documentclass[preprintnumbers,twocolumn,amsmath,amssymb,pre]{revtex4}

\usepackage{graphicx}

\begin{document}

\title{Spontaneous dissipation of elastic energy by self-localizing thermal runaway}

\author{S. Braeck}
\email{Simen.Brack@iu.hio.no}
\affiliation{Faculty of Engineering, Oslo University College, Postbox 4 St. Olavs Plass, N-0130 Oslo, Norway}
\author{Y.Y. Podladchikov}
\author{S. Medvedev}
\affiliation{Physics of Geological Processes (PGP), University of Oslo, P.O. Box 1048 Blindern,
N-0316 Oslo, Norway}

\begin{abstract}

Thermal runaway instability induced by material softening due to shear heating represents a potential mechanism for mechanical
failure of viscoelastic solids. In this work we present a model based on a continuum formulation of a viscoelastic material
with Arrhenius dependence of viscosity on temperature, and investigate the behavior of the thermal runaway phenomenon by
analytical and numerical methods. Approximate analytical descriptions of the problem reveal that onset of thermal runaway
instability is controlled by only two dimensionless combinations of physical parameters. Numerical simulations of the
model independently verify these analytical results and allow a quantitative examination of the complete time evolutions
of the shear stress and the spatial distributions of temperature and displacement during runaway instability. Thus we
find that thermal runaway processes may well develop under nonadiabatic conditions. Moreover, nonadiabaticity of
the unstable runaway mode leads to continuous and extreme localization of the strain and temperature profiles in space,
demonstrating that the thermal runaway process can cause shear banding. Examples of time evolutions of the spatial
distribution of the shear displacement between the interior of the shear band and the essentially nondeforming material
outside are presented. Finally, a simple relation between evolution of shear stress, displacement, shear-band width and
temperature rise during runaway instability is given.

\end{abstract}

\maketitle

\section{Introduction\label{sec:intro}}

Many materials are known to exhibit fracture phenomena that, in apparent contradiction to the expected failure behavior usually
associated with plastic yielding or brittle cracking, are characterized by shear deformation localized along a single or a few
bands. Ordinarily, one would expect brittle fracture to occur by the opening of microscopic cracks along the fault plane, while
conventional plastic yielding associated with deformation of crystals is known to be pervasive and involve work-hardening which
seem to be incompatible with the localization of the deformation in shear bands indicating some form of work-softening. This
distinct mode of shear failure is observed in a variety of materials, including some amorphous solids such as
polymers~\cite{Lu,Li,Cross} and bulk metallic glasses~\cite{Greer,Wright,Lewandowski,Hays}, rocks under high confining
pressure~\cite{Obata,Renshaw,Wiens} and crystalline solids deformed rapidly in impact experiments~\cite{Miller,TWWright}.

The formation of shear bands in bulk metallic glasses and rocks under high confinement have attracted much attention because,
in these materials, the mechanism responsible for the weakening of the material in the bands often initiate instabilities that
lead to catastrophic shear failure along one dominant band. Accordingly, shear banding represents the primary mode of failure
in many of these material systems. Experimental studies of fracture surfaces of rock samples subjected to high confining pressures
demonstrate that the region near the dominant band is essentially devoid of microcracks~\cite{Renshaw,Schulson}, pointing towards a
mechanism leading to viscous sliding without cracking. Similarly, studies of fracture surfaces of metallic glass samples suggest
that catastrophic failure is attributable to a large decrease of the viscosity of the material in the catastrophic
shear band~\cite{Wright,Pampillo}. As a consequence, these materials appear to fail in a globally brittle, but locally ductile,
manner. The shear bands in metallic glasses, 10--20 nm thick~\cite{Pekarskaya}, seem to be accompanied by significant local
increases in temperature. Vein patterns and solidified drops on the fracture surfaces~\cite{Wright,Hays,Bengus} indicative of
melting of material during catastrophic failure as well as indirect experimental measurements of temperature rises during
shear-band operation~\cite{Lewandowski} lend support to this view. Similarly, localized shear failure occurring in the deeper
parts of the Earth's lithosphere, believed to be related to earthquakes located several tens of kilometers below the Earth's
surface, may involve substantial rises in temperature. Geological field observations of melted rock in the form of cm thick
pseudotachylyte layers along shear faults provide evidence for considerable heat dissipation during such
failure~\cite{Obata,Torgeir,Haakon,Timm}. The catastrophic shear failure process thus seems to be characterized by
spontaneous release of a substantial proportion of the stored elastic energy as heat in the region of the
rapidly forming shear band.

Apart from the observed differences in the qualitative macroscopic features of localized shear failure as compared to
ordinary brittle and plastic failure behaviors, the circumstances under which shear failure seem to occur indicate that this mode
of failure may not be ascribed to the conventional mechanisms of opening of microscopic cracks or crystallographic slip by
dislocation motion. For instance, the closure of cracks at high confining pressures, non-planar crystal structure of minerals
and disorder of mineral grain orientations are all factors that inhibit these mechanisms from operating in rocks
in the Earth's interior. In metallic glasses the high degree of structural disorder cause dislocations to experience a large
number of obstacles, reducing their mobility and inhibiting plastic flow. Because of the absence of these basic weakening
mechanisms, mantle rocks and bulk metallic glasses can demonstrate exceptionally high strengths. However, although these materials
have strengths approaching the theoretical shear strength limit at which atomic bonds break~\cite{Frenkel}, the fact that
shear-band thicknesses are usually much larger than interatomic spacing suggests the existence of alternative mechanisms
responsible for the catastrophic shear failure. In metallic glasses it has been proposed that granular structure on the microscopic
scale may be largely involved in determining the finite width of a shear band~\cite{Zhang}. Discrete element modeling have shown that
granular packings subjected to compression may fail by the formation of shear bands or faults due to dilatancy, predicting a
shear-band width of about ten grain diameters~\cite{Astrom,Francois}. However, experimental investigations of deformation in a class
of shear yielding polymers~\cite{Lu} show that, despite of the fact that the molecular and microstructural deformation mechanisms are
rather different, the phenomenon of shear banding in these materials is strikingly similar to the localization of deformation in
metallic materials. This suggests that the general large-scale features of the shear banding phenomenon might be appropriately modeled
using a continuum formulation as long as the grain size is small.

Several earlier works have investigated the possibility of shear
failure induced by softening mechanisms facilitating ductile or
plastic-like deformation. Mainly two explanations have been proposed
for the observed localization of
shear~\cite{Pampillo,Molinari,Johnson,Schuh}. The first, based on
various micromechanical theories developed by Spaepen, Argon, Falk
and Langer, and others in order to describe plasticity in amorphous
materials~\cite{Zhang,Johnson,Spaepen,Argon,ArgonShi,FalkLanger,Langer,Greer09},
suggests that material softening due to structural changes is a
mechanism for strain localization. In this case it is usually
assumed that the local heat generation during deformation is only a
secondary effect and is important to the evolution of the material
in the bands only in the later stages of slip. In a recent work,
however, Manning, Langer and Carlson~\cite{Manning} proposed that
the heat generated by plastic deformation is dissipated in the
system's configurational degrees of freedom and raises an effective
temperature rather than the usual kinetic temperature. Thus it was
shown that the effective temperature could provide a mechanism for
strain localization.

The second explanation, first proposed by Griggs and
Baker~\cite{Griggs} and later developed by Ogawa~\cite{Ogawa} in
order to explain the occurrence of deep-focus earthquakes,
introduces the concept of thermal softening according to which the
material is weakened primarily due to the effect of local heating.
Local heating increases the temperature which leads to a
corresponding decrease in the strongly temperature dependent
viscosity. Recent theoretical results~\cite{BP} have demonstrated
that, even under nonadiabatic conditions, the thermal softening
mechanism may induce a thermal runaway instability exhibiting
progressive strain localization, thus leading to shear-band
formation and consequent material failure. These results are
apparently consistent with experimental results on bulk metallic
glasses~\cite{Lewandowski,Zhurkin} showing that shear-band operation
cannot be fully adiabatic.

In the present work we expand on the theoretical investigations of thermal runaway instability in solids, already
presented in abbreviated form in Ref.~[\onlinecite{BP}]. Our model is based on a simple continuum formulation of a viscoelastic
medium, i.e., the rheology contains both viscous and elastic components. The viscous material response is supposed to be induced
by thermally activated processes, yielding a strongly temperature dependent viscosity. Thus the model accounts for nonelastic
mechanical responses shown by real materials even below the conventional elastic limit, such as the well-known phenomena of creep
or relaxation. The temperature in the system is given by the equation for energy conservation. The viscoelastic rheology equation,
governing the mechanical behavior of the material, is then coupled to the energy conservation equation through temperature dependent
viscosity. We shall make the assumption that initiation of localized deformation is triggered by small (but macroscopic) local
heterogeneities or thermal fluctuations in the otherwise large-scale homogeneous and isotropic material. Hence, an increase in
strain rate in a weaker zone may cause a local temperature rise due to viscous dissipation, which weakens the zone even further.
As a consequence, the local increase in strain rate and temperature may amplify strongly because of the effect of shear
heating-induced thermal softening of the material. Accordingly, catastrophic shear failure may occur as a result of thermal runaway
instability.

The paper is organized as follows. In Sec.~\ref{sec:model} a simple viscoelastic model is introduced and the basic governing
equations are formulated. In Sec.~\ref{sec:anan} analytical methods are used to derive the condition for thermal runaway to occur
in the adiabatic limit and to estimate the resulting adiabatic temperature rise. Sec.~\ref{sec:trconditions} presents a linear
analysis for the purpose of determining the conditions necessary for thermal runaway to occur for the general case by taking into
account the effects of thermal conduction. Numerical solutions to the exact equations are presented in
Sec.~\ref{sec:numapp}, allowing us to quantify the later stages of the thermal runaway process and, in particular, the effects
of thermal diffusion will be addressed. We proceed in Sec.~\ref{sec:stressdrop} to derive analytically a relatively simple
relation between the evolution of stress, displacement and temperature rise inside the shear band for an adiabatic runaway
process. The accuracy of this relation is then evaluated by comparing it to the numerical results for nonadiabatic processes.
Finally, we summarize the main conclusions in Sec~\ref{sec:conc}.

\section{Model\label{sec:model}}

\begin{figure}[tbp]
\begin{center}
\includegraphics[width=8cm]{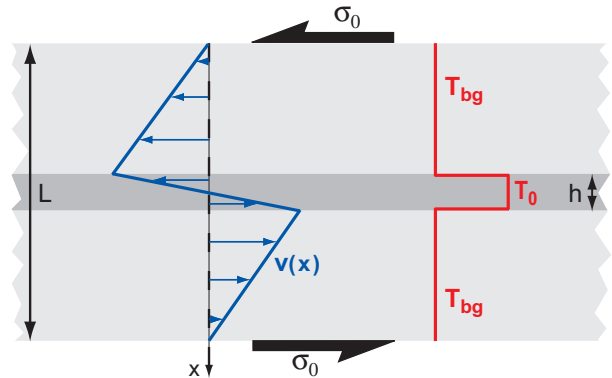}
\caption{(Color online) Initial setup of the viscoelastic slab model
discussed in the text (cross-section in the $xy$-plane). The slab is
in a state of stress of simple shear with zero velocity ($v=0$)
boundary conditions. The shear stress $\sigma$, constant throughout
the slab, initially ($t=0$) equals the value $\sigma_0$. The lines
show the initial velocity, $v(x)$, and temperature, $T(x)$,
profiles. The shaded region illustrates a small perturbation in
temperature $T=T_0$ of width $h$ at the slab center. Elsewhere, the
background temperature is $T=T_{bg}$. The geometry of the strain
rate profile concurs with that of the temperature profile.}
\label{fig:modelsetup}
\end{center}
\end{figure}

We consider a model consisting of an infinite viscoelastic slab having a finite width $L$ in the $x$-direction;
i.e., the geometry is that of a solid bounded by a pair of parallel infinite planes (see Fig.~\ref{fig:modelsetup}).
We assume that the slab is in a condition of simple shear such that the only nonzero component of the displacement field is
the y-component, which we denote by $u$. Then the shear stress $\sigma_{xy}(=\sigma_{yx})$, hereafter denoted by $\sigma$, is
constant throughout the slab and hence only a function of the time $t$ (see Eq.~(\ref{eq:scndgoveq}) below). Our purpose is
to examine spontaneous modes of internal failure in the slab. Therefore, in order to eliminate any additional effects of
far-field deformation that could either aid or trigger failure, we impose zero velocity at the slab's boundaries while we
assume that the initial shear stress in the slab is $\sigma_0$. It is thus assumed that the shear stress ${\sigma}$ has
attained a value $\sigma_0$ at $t=0$ regardless of the slab's loading history which is not considered in the present model.
Accordingly, the shear stress in the slab decreases with time from it's maximum value $\sigma_0$ due to relaxation and viscous
deformation in the interior. Our particular model setup with zero velocity boundary conditions amounts to searching for the
ultimate conditions for which the slab will fail, that is, if it fails at these conditions it is expected to fail earlier at
any others. The temperature $T$ in the slab initially equals a background temperature $T_{bg}$ except in the small central
region having width $h$ and a slightly elevated temperature $T_0$. The small thermal perturbation introduced in the central
region ensures that initiation of localized deformation occurs in the neighborhood of the slab's center. The
boundaries~($x=\pm L/2$) are maintained at the temperature $T_{bg}$.

Assuming inertial effects are negligible, it follows from the translational symmetry in the y- and z-directions of the present model
that the shear stress must satisfy the reduced equation for conservation of momentum
\beq
\frac{\partial \sigma}{\partial x} = 0\,,
\label{eq:scndgoveq}
\eeq
showing that $\sigma$ is independent of $x$. Without loss of generality the remaining components of the stress tensor are regarded as
zero. The viscoelastic rheology is represented by the Maxwell model~\cite{Malvern} for which the strain rate
is given by
\beq
\frac{\partial v}{\partial x} = \frac{1}{\mu(T,\sigma)}\sigma + \frac{1}{G}\frac{\partial\sigma}{\partial t}\,.
\label{eq:Maxwellrheology}
\eeq
Here $v(x,t)$ is the velocity in the y-direction, $G$ is the constant shear modulus and $\mu(T,\sigma)$ is the viscosity.
The first and the last term on the r.h.s. of the equation represent the viscous and elastic components of the
strain rate, respectively. Assuming that the viscosity of the material is governed by thermally activated processes, the
functional dependence of the viscosity on temperature and shear stress may be approximated as
\beq
\mu(T,\sigma)=A^{-1}e^{E/RT}\sigma^{1-n}\, .
\label{eq:viscosity}
\eeq
Here $A^{-1}$ is a pre-exponential constant, $n$ is a constant characterizing the dominant creep mechanism,
$E$ is the activation energy of creep and \mbox{$R=8.3$ JK$^{-1}$mole$^{-1}$} is the universal gas constant.
Thus the viscosity has Arrhenius dependence on temperature and it is, in general, a non-linear
function of the shear stress.

We simplify the mathematical problem by integration of
Eq.~(\ref{eq:Maxwellrheology}) to eliminate the velocity from the
system of equations. Utilizing the zero velocity boundary
conditions, this gives \beq
\int_{-\frac{L}{2}}^{\frac{L}{2}}Ae^{-\frac{E}{RT}}\sigma^n +
\frac{1}{G}\frac{\partial\sigma}{\partial t}dx
=\int_{-\frac{L}{2}}^{\frac{L}{2}}\frac{\partial v}{\partial x}dx=0
\,, \label{eq:bcondint} \eeq and we thus obtain the equation which
governs the time-dependence of $\sigma$: \beq \frac{\partial
\sigma}{\partial t} &=& -\frac{GA}{L}\sigma^n
\int_{-\frac{L}{2}}^{\frac{L}{2}}e^{-\frac{E}{RT}}dx\, .
\label{eq:finstresseq} \eeq The temperature is determined by the
equation for energy conservation \beq \frac{\partial T}{\partial t}
&=& \kappa \frac{\partial^2 T}{\partial x^2} + \frac{1}{C}\sigma
\left( \frac{\partial v}{\partial
x}-\frac{1}{G}\frac{\partial\sigma}{\partial t} \right)\, ,
\label{eq:fintempeqone} \eeq where $\kappa$ is the thermal
diffusivity and $C$ denotes the heat capacity per volume. The last
term in Eq.~(\ref{eq:fintempeqone}) accounts for dissipation in the
system and thus includes only the viscous part of the strain rate.
Upon substituting for $\partial v/\partial x$ the expression in
Eq.~(\ref{eq:Maxwellrheology}), the energy equation becomes \beq
\frac{\partial T}{\partial t} &=& \kappa \frac{\partial^2
T}{\partial x^2} + \frac{A}{C}\sigma^{n+1}e^{-\frac{E}{RT}}\, .
\label{eq:fintempeqtwo} \eeq Equations (\ref{eq:finstresseq}) and
(\ref{eq:fintempeqtwo}) with the specified initial and boundary
conditions constitute a closed set of equations for $T(x,t)$ and
$\sigma(t)$. These two equations provide the mathematical basis for
calculating all the quantities of interest during the deformation
process since, from the solutions to these equations for $T$ and
$\sigma$, one may calculate the strain rate directly from
Eq.~(\ref{eq:Maxwellrheology}). We note that, because $\sigma(t)$ is
independent of $x$, it follows from Eq.~(\ref{eq:Maxwellrheology})
that the geometry of the strain rate profile at any instant concurs
with that of the temperature profile $T(x,t)$.

From the considerations above it is clear that we can infer
important information about the deformation processes by studying
the temperature rises in the system. Useful analytical results can
be obtained by invoking approximations appropriate for describing
the initial stages of evolutions of $T$ and $\sigma$ for which
temperature rises are comparatively small. If instabilities develop,
substantial temperature rises are possible, but correspondingly
large decreases in shear stress act against unlimited growth.
Therefore, to quantify the later stages of the thermal runaway
processes, we shall take the maximum temperature rise, defined as
\beq \Delta T_{max}= T_{max}-T_0\,, \label{eq:defmaxtemp} \eeq where
$T_{max}$ is the maximum temperature with respect to both time and
position, as the appropriate physical quantity to study. We solve
the system of equations~(\ref{eq:finstresseq}) and
(\ref{eq:fintempeqtwo}) and estimate $\Delta T_{max}$ using
approximate analytical~(Sec.~\ref{sec:anan} and
\ref{sec:trconditions}) and numerical~(Sec.~\ref{sec:numapp})
methods.

\section{Adiabatic case\label{sec:anan}}

Important insight into the viscoelastic model is gained by first examining the limiting case
of adiabatic heating for which the first term on the r.h.s. of Eq.~(\ref{eq:fintempeqtwo}) is neglected. Then the
temperature is determined by the reduced equation
\beq
\frac{\partial T}{\partial t} = \frac{A}{C}\sigma^{n+1}e^{-\frac{E}{RT}}\, .
\label{eq:adiabatictempeq}
\eeq

\subsection{Analytical solution\label{subsec:adtrconditions}}

As aid towards understanding the slab's deformation behavior, we first examine the time evolutions of temperature and
shear stress in the initial stages for which temperature changes may be considered comparatively small. In this limit it is
possible to obtain approximate analytical solutions to elucidate the deformation problem.

As a first approximation, we insert the initial condition for $T$ in Eq.~(\ref{eq:finstresseq}) and perform the
integration over space to obtain
\beq
\frac{\partial\sigma}{\partial t} = -GAe^{-\frac{E}{RT_0}}\Delta_p\sigma^n \,,
\label{eq:initsigma}
\eeq
where
\beq
\Delta_p=\frac{h}{L} + \left( 1-\frac{h}{L} \right) \frac{\mu(T_0,\sigma_0)}{\mu(T_{bg},\sigma_0)}
\label{eq:initpertfact}
\eeq
is a factor which characterizes the initial perturbation. At this point,
it is convenient to introduce the quantity $\mu_0\equiv\mu(T_0,\sigma_0)$ and the dimensionless stress and time
\beq
\tilde{\sigma}=\frac{\sigma}{\sigma_0}\,,\ \ \tilde{t}=\frac{t}{\tau_r}\,,
\label{eq:dimlessst}
\eeq
with
\beq
\tau_r=\frac{\mu_0}{2G\Delta_p}
\label{eq:relaxtime}
\eeq
denoting a characteristic time for stress relaxation in the system. The simplified dimensionless form of the
equation is thus
\beq
\frac{\partial\tilde{\sigma}}{\partial\tilde{t}}=-\frac{1}{2}\tilde{\sigma}^n\,.
\label{eq:dimlessadstresseq}
\eeq
This equation can be integrated directly by separation of variables and, for the initial condition
$\tilde{\sigma}=1$, the corresponding solutions are
\beq
\tilde{\sigma}=e^{-\frac{1}{2}\tilde{t}}\,,\ \ \ \ n=1\,,
\label{eq:adstresssolno}
\eeq
and
\beq
\tilde{\sigma}=\left[ \frac{n-1}{2}\tilde{t}+1 \right]^{\frac{1}{1-n}}\,,\ \ \ \ n>1\,.
\label{eq:adstresssolng}
\eeq
Another simplification follows from Taylor expanding $1/T$ to first order about the initial temperature $T_0$
in the central perturbed zone.
Using this approximation, the exponential function in Eq.~(\ref{eq:adiabatictempeq}) may be written
\beq
e^{-\frac{E}{RT}}\approx e^{-\frac{E}{RT_0}}e^{\frac{E(T-T_0)}{RT_0^2}}\,.
\label{eq:tempexpansion}
\eeq
Defining the dimensionless temperature
\beq
\theta=\frac{E(T-T_0)}{RT_0^2}\,
\label{eq:dimlesstemp}
\eeq
and using Eqs.~(\ref{eq:dimlessst}) and (\ref{eq:tempexpansion}), we can rewrite Eq.~(\ref{eq:adiabatictempeq}) on
dimensionless form as
\beq
\frac{\partial\theta}{\partial\tilde{t}}=\left( \frac{\sigma_0}{\sigma_c} \right)^2\tilde{\sigma}^{n+1}e^{\theta}\,,
\label{eq:dimlesstempeq}
\eeq
which must satisfy the initial condition $\theta=0$ in the perturbed zone. Here we have introduced the new quantity
\beq
\sigma_c=\sqrt{2\Delta_p\frac{GCR}{E}}T_0\,,
\label{eq:criticalstress}
\eeq
having the same dimension as stress. Now, by substituting the solutions~(\ref{eq:adstresssolno}) and (\ref{eq:adstresssolng}) for
$\tilde{\sigma}$ in Eq.~(\ref{eq:dimlesstempeq}), and once again integrating by separation of variables, we find the
solutions
\beq
\theta = -\ln \left[ \left( \frac{\sigma_0}{\sigma_c} \right)^2 \left( e^{-\tilde{t}}-1 \right) + 1 \right] \,, \ \ \ \ n=1\,,
\label{eq:adtempsolno}
\eeq
and
\beq
\theta =
\mbox{} - \ln\left\{ \left( \frac{\sigma_0}{\sigma_c} \right)^2 \left[ \left( \frac{n-1}{2}\tilde{t} + 1 \right)^{\frac{2}{1-n}}
-1 \right] + 1 \right\}\,, \nonumber \\
n>1\,. \nonumber \\
\label{eq:adtempsolng}
\eeq
Both solutions~(\ref{eq:adtempsolno}) and (\ref{eq:adtempsolng}) are seen to exhibit two distinct modes of evolution, depending on
the value of $\sigma_0$. Indeed, the solutions are bounded only if $\sigma_0<\sigma_c$. If this condition is violated, the temperature
grows unlimited at the critical times
\beq
\tilde{t}_{cr} = -\ln\left[ 1-\left( \frac{\sigma_c}{\sigma_0} \right)^2 \right]\,, \ \ \ \ n=1\,,
\label{eq:crittimeno}
\eeq
and
\beq
\tilde{t}_{cr} = \frac{2}{n-1} \left\{ \left[ 1-\left( \frac{\sigma_c}{\sigma_0} \right)^2 \right]^{\frac{1-n}{2}}
-1 \right\} \,, \ \ \ \ n>1\,.
\label{eq:crittimeng}
\eeq
The dramatic change in growth of temperature as $\sigma_0$ exceeds $\sigma_c$ indicates that, under adiabatic conditions,
$\sigma_c$ plays the role of a critical stress above which thermal runaway occurs.

A few remarks concerning the validity of Eqs.~(\ref{eq:adtempsolno})--(\ref{eq:crittimeng}) are appropriate here. The main effects of
the approximations adopted in deriving these equations is that the heat production rate becomes too large for large temperature
rises and that temperatures may increase without limit. In contrast, the exact equations always give finite values for temperature
rises. Nevertheless, solutions~(\ref{eq:adtempsolno}) and (\ref{eq:adtempsolng}) are expected to yield reasonable approximations at
least into the very early stages of development of adiabatic thermal runaway. Hence, we also expect that these simplified estimates
correctly predict the condition for adiabatic thermal runaway to occur. That this indeed is the case, will be independently verified
by numerical methods in section~\ref{sec:numapp}.

\subsection{Maximum temperature rise during adiabatic thermal runaway\label{subsec:adtemp}}

We now discuss the later stages of adiabatic thermal runaway, for which the analytical solutions obtained in the previous section are
inapplicable. Then, a simple analytical estimate of the maximum temperature rise (Eq.~(\ref{eq:defmaxtemp})) can be made by considering
overall energy balance as follows.

Integration of Eq.~(\ref{eq:adiabatictempeq}) over space gives
\beq
\int_{-\frac{L}{2}}^{\frac{L}{2}}\frac{\partial T}{\partial t}dx
= \frac{A}{C}\sigma^{n+1}\int_{-\frac{L}{2}}^{\frac{L}{2}}e^{-\frac{E}{RT}}dx \,.
\label{eq:inttempeq}
\eeq
The expression on the r.h.s. can be obtained from Eq.~(\ref{eq:finstresseq}), giving
\beq
\frac{A}{C}\sigma^{n+1}\int_{-\frac{L}{2}}^{\frac{L}{2}}e^{-\frac{E}{RT}}dx
= -\frac{L}{GC}\sigma\frac{\partial\sigma}{\partial t}\,.
\label{eq:multipstresseq}
\eeq
Combining Eqs.~ (\ref{eq:inttempeq}) and (\ref{eq:multipstresseq}) we obtain
\beq
\frac{\partial}{\partial t} \left[ \int_{-\frac{L}{2}}^{\frac{L}{2}}Tdx + \frac{L}{2GC}\sigma^2 \right] = 0\,,
\label{eq:combeq}
\eeq
which may be integrated over time to give
\beq
\int_{-\frac{L}{2}}^{\frac{L}{2}}\Delta T(x,t)dx = \frac{L}{C}\frac{\sigma_0^2-\sigma(t)^2}{2G}\,,
\label{eq:timeintcombeq}
\eeq
where $\Delta T(x,t)=T(x,t)-T(x,0)$. Now, it is reasonable to assume that the viscous part of the deformation mainly occurs
within the initially perturbed zone $|x|\leq h/2$. Then, essentially no heat will be dissipated outside this zone, and
because diffusion of heat is assumed negligible for the present case, the temperature rises in these outer regions become
vanishingly small. Inside the perturbed region, dissipation of heat is expected to be distributed monotonously. To understand
the reason, recall that in this region the initial temperature, and therefore the initial viscosity, are uniformly distributed.
Accordingly, as long as the process is adiabatic, the temperature rise in the same region will be approximately uniform, i.e.,
independent of $x$. As a result, the integral in Eq.~(\ref{eq:timeintcombeq}) may be replaced by a simple integration of uniform
temperature rise over the region $|x|\leq h/2$. Performing the integration, we find that the adiabatic maximum temperature rise
inside the perturbed region is given by
\beq
\Delta T_{max}^a = \frac{\sigma_0^2L}{2GCh}.
\label{eq:maxadtemprise}
\eeq
In writing Eq.~(\ref{eq:maxadtemprise}) we have made the assumption that the runaway process continues until the stress
$\sigma(t)$ in the slab is much smaller than $\sigma_0$.
We recognize the quantity $\sigma_0^2/2G$ as the elastic energy per unit volume stored in the slab in its initial state.
Eq.~(\ref{eq:maxadtemprise}) reflects the fact that during thermal runaway, and under the assumptions made, all the elastic energy
in the system spontaneously dissipates \emph{uniformly} as heat in the initially perturbed zone. As was mentioned earlier, the geometry
of the strain rate profile concurs at any instant with that of the temperature profile. An immediate consequence of the adiabatic assumption
is therefore that the width of the shear band formed during runaway equals the width of the initially perturbed zone in which the uniform
temperature rise occurs. Although the adiabatic approximation provides a simple means to obtain important insight into the process, neglect
of thermal diffusion may give misleading results. The extent to which the adiabatic assumption is valid will be investigated by numerical
methods in Sec.~\ref{sec:numapp}.

\section{Conditions for thermal runaway. The general case\label{sec:trconditions}}

In this section we attempt to determine the conditions necessary for thermal runaway to occur for the general case by taking into
account the effects of thermal conduction. For that purpose, we investigate the stability of the coupled equations
(\ref{eq:finstresseq}) and (\ref{eq:fintempeqtwo}) by a linear analysis. Then, as an approximation
for the initial stages, we insert the initial conditions for $\sigma$ and $T$ in Eq.~(\ref{eq:finstresseq}) and
carry out the integration over space to obtain the simplified equation
\beq
\frac{\partial\sigma}{\partial t} = -\frac{\sigma_0}{2\tau_r}\, ,
\label{eq:linstressone}
\eeq
which has the solution
\beq
\sigma(t)=\sigma_0 \left[ 1-\frac{t}{2\tau_r} \right] \,.
\label{eq:linstresstwo}
\eeq
Here $\tau_r$ is the relaxation time given by eq.~(\ref{eq:relaxtime}).
Using this expression, we expand $\sigma^{n+1}$ to first order in time, yielding
\beq
\sigma(t)^{n+1}\approx\sigma_0^{n+1} \left[ 1-(n+1)\frac{t}{2\tau_r} \right]\, .
\label{eq:linstressthree}
\eeq
The expansion is valid for small times $t\ll\tau_r$.
Equation~(\ref{eq:fintempeqtwo}), determining the temperature rise in the system, may now be approximated
by substituting for $\sigma^{n+1}$ the result obtained in Eq.~(\ref{eq:linstressthree}):
\beq
\frac{\partial T}{\partial t} = \kappa\frac{\partial^2 T}{\partial x^2} + \frac{\sigma_0^2}{C\mu_0} \left[ 1-(n+1)\frac{t}{2\tau_r} \right]
e^{\frac{E}{R} \left( \frac{1}{T_0}-\frac{1}{T} \right) }\, .
\label{eq:lintempone}
\eeq
Our aim is to obtain the conditions for which the perturbation in the central region becomes unstable. Hence,
in the following discussion we consider only the temperature inside the central region $|x|\leq h/2$.
Assume that the temperature evolution is continuous and smooth and that there exists a steady temperature $T_{ss}$
for which $\partial T_{ss}/\partial t = 0$. Then $T_{ss}$ satisfies the reduced equation for steady flow
\beq
\kappa\frac{\partial^2 T_{ss}}{\partial x^2} = -\frac{\sigma_0^2}{C\mu_0} \left[ 1-(n+1)\frac{t}{2\tau_r} \right]
e^{\frac{E}{R} \left( \frac{1}{T_0}-\frac{1}{T_{ss}} \right) }\, .
\label{eq:steadystate}
\eeq
The stability of the steady state solution may be investigated by superposing a perturbation
$\delta$ on $T_{ss}$. The resulting non-steady temperature $T=T_{ss}+\delta$ must satisfy
Eq.~(\ref{eq:lintempone}), describing time-dependent flow. For simplicity, we restrict the analysis
to arbitrarily small $\delta$. In this case, since $\delta/T_{ss}\ll 1$, we have
$1/T\approx 1/T_{ss}-\delta/T_{ss}^2$, and we obtain for the exponential term in Eq.~(\ref{eq:lintempone})
\beq
e^{\frac{E}{R} \left( \frac{1}{T_0}-\frac{1}{T} \right) } &\approx& e^{\frac{E}{R} \left( \frac{1}{T_0}-\frac{1}{T_{ss}} \right) }
e^{\frac{E\delta}{RT_{ss}^2}} \nonumber \\
&\approx& e^{\frac{E}{R} \left( \frac{1}{T_0}-\frac{1}{T_{ss}} \right) } \left( 1-\frac{E\delta}{RT_{ss}^2} \right) \, ,
\label{eq:linexponential}
\eeq
where we keep only the leading term in the last expansion because $E\delta/RT_{ss}^2\ll 1$ for arbitrarily small
$\delta$. Using this expansion and utilizing Eq.~(\ref{eq:steadystate}), Eq.~(\ref{eq:lintempone})
reduces to a linearized equation for $\delta$:
\beq
\frac{\partial\delta}{\partial t} &=& \kappa\frac{\partial^2\delta}{\partial x^2} \nonumber \\
& & + \frac{\sigma_0^2E}{C\mu_0RT_{ss}^2}e^{\frac{E}{R} \left( \frac{1}{T_0}-\frac{1}{T_{ss}} \right) }
\left( 1-(n+1)\frac{t}{2\tau_r} \right) \delta \, . \nonumber \\
\label{eq:lindeltaone}
\eeq
In our search for spontaneous thermal runaway modes we are interested in finding the conditions for which
the initial perturbation $T_0$ instantly starts to increase. If we now anticipate that there is a sharp
transition boundary between the stable modes, where $T_0$ instantly drops, and the unstable modes, where
$T_0$ instantly grows, one may expect that $T_0$ plays the role of a steady state right at the transition
boundary. Accordingly, to obtain the conditions for spontaneous runaway modes, one should analyze the case
in which the initial perturbation $T_0$ itself becomes a steady state and therefore choose $T_{ss}\approx T_0$
in Eq.~(\ref{eq:lindeltaone}). This yields the equation
\beq
\frac{\partial\delta}{\partial t} = \kappa\frac{\partial^2\delta}{\partial x^2}
+ \frac{\sigma_0^2E}{C\mu_0RT_0^2} \left( 1-(n+1)\frac{t}{2\tau_r} \right) \delta \, .
\label{eq:lindeltatwo}
\eeq
Then, normalizing time by the relaxation time given by Eq.~(\ref{eq:relaxtime}), and introducing new dimensionless variables
\beq
\tilde{\delta}=\frac{\delta}{E/R}\,,\ \ \tilde{x}=\frac{x}{h}\,,
\label{eq:dimlessvarstab}
\eeq
we recast Eq.~(\ref{eq:lindeltatwo}) into dimensionless form as
\beq
\frac{\partial\tilde{\delta}}{\partial\tilde{t}} =
\frac{\tau_r}{\tau_d}\frac{\partial^2\tilde{\delta}}{\partial\tilde{x}^2}
+ \left( \frac{\sigma_0}{\sigma_c} \right)^2 \left( 1-(n+1)\frac{\tilde t}{2} \right) \tilde{\delta}\, .
\label{eq:dimlessstabeq}
\eeq
Here $\tau_d=h^2/\kappa$ is the characteristic thermal diffusion time for diffusion of heat away from the central
region, and $\sigma_c$ is the characteristic stress introduced earlier in~(\ref{eq:criticalstress}).
Note that two combinations of physical parameters appear in this simplified equation, namely
$\sigma_0/\sigma_c$ and  $\tau_r/\tau_d$. The first combination entered the analysis in Sec.~\ref{subsec:adtrconditions},
in which it was interpreted as the factor controlling the stability of the system in the adiabatic limit.
The second combination, which is the ratio of the relaxation time to the thermal diffusion time, will modify the stability
criterion for processes which are not adiabatic. The effect of this dimensionless variable on the conditions for thermal
runaway will be addressed in the analysis below.

\begin{figure}[tbp]
\begin{center}
\includegraphics[width=8cm]{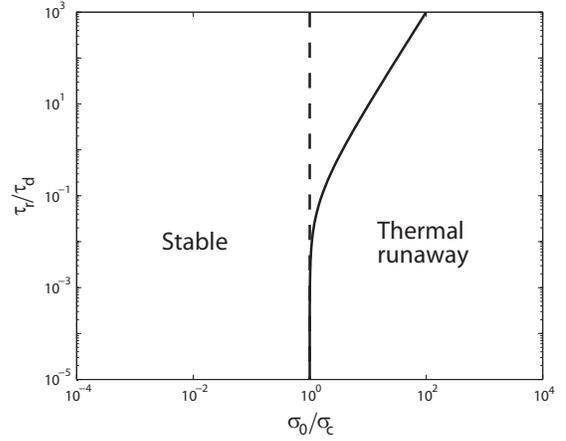}
\caption{Plots of the critical values of $\sigma_0/\sigma_c$
defining the transition boundaries between stable deformation modes
and unstable thermal runaway modes according to approximate
theoretical predictions. The solid curve shows the results of the
linear stability analysis taking thermal conduction into account, as
given by the expression in Eq.~(\ref{eq:thrcond}). The critical
values of $\sigma_0/\sigma_c$ are seen to increase with increasing
$\tau_r/\tau_d$. The dashed line shows the results of the adiabatic
analysis made in Sec.~\ref{sec:anan} which completely neglects the
effects of thermal conduction and thus predicts a threshold stress
which is independent of $\tau_r/\tau_d$.} \label{fig:anstabsol}
\end{center}
\end{figure}

Assuming, for simplicity, that $\tilde{\delta}$ vanish at the ``boundaries'' $\tilde{x}=\pm 1/2$
and that $\tilde{\delta}$ is a positive, even function, the general solution to Eq.~(\ref{eq:dimlessstabeq})
can be represented as a sum of particular solutions as
\beq
\tilde{\delta} = \sum_{m=0}^{\infty}f_m \left( \tilde{x},\tilde{t} \right) \, ,
\label{eq:Fouriersol}
\eeq
with
\beq
f_m \left( \tilde{x},\tilde{t} \right) &=& B_me^{\left[ -\frac{\tau_r}{\tau_d}k_m^2
+ \left( \frac{\sigma_0}{\sigma_c} \right)^2 \right] \tilde{t}
- \left( \frac{\sigma_0}{\sigma_c} \right)^2 \frac{(n+1)\tilde{t}^2}{4}} \nonumber \\
& & \times\cos \left( k_m\tilde{x} \right) \,. \nonumber \\
\label{eq:particsol}
\eeq
Here $k_m=(2m+1)\pi$ denote the frequencies of the perturbation satisfying the required boundary
conditions, and $B_m$ denote the amplitudes. It follows from the general formula
for Fourier cosine coefficients that $B_m$ must decrease with increasing $m$. Furthermore, we note that $k_m$ increases with
increasing $m$. Accordingly, $f_0$ is the largest term in the series expansion, and it dominates
the stability of the perturbation in~(\ref{eq:Fouriersol}). To proceed, we must analyze the time evolution of the perturbation,
which is controlled by the argument of the exponential in $f_0$. For the case when the coefficient in the leading term in the argument
is negative or zero, $f_0$ becomes a monotonically decreasing function of $\tilde{t}$. As a result, $\tilde{\delta}$ decreases with time,
implying a stable situation. In the opposite case, when the coefficient in the leading term of the argument is positive, $f_0$ increases
with time until it reaches a maximum for which $\partial f_0/\partial\tilde{t}=0$. Solving this equation, we obtain the characteristic time
\beq
\tilde{t}_{max} = \frac{2}{n+1} \left[ 1 - \left( \frac{\sigma_0}{\sigma_c} \right)^{-2} \frac{\tau_r}{\tau_d}\pi^2 \right]
\label{eq:chtimegrowth}
\eeq
above which the perturbation begins to decrease, and the corresponding maximum of $f_0$:
\beq
f_0 \left( \tilde{x}, \tilde{t}_{max} \right) = B_0e^{\frac{1}{n+1} \left( \frac{\sigma_0}{\sigma_c} \right)^{-2}
\left[ \left( \frac{\sigma_0}{\sigma_c} \right)^2 - \frac{\tau_r}{\tau_d}\pi^2 \right]^2}
\cos \left( \pi\tilde{x} \right) \, . \nonumber \\
\label{eq:fmax}
\eeq
In Eqs.~(\ref{eq:chtimegrowth}) and (\ref{eq:fmax}) we used that $k_0=\pi$.
Thus, for certain conditions, the solution to this linearized perturbed problem predicts only a limited increase in temperature. Due to
the linear approximation of the exponential functions in the linear analysis, however, heat is produced at
a rate which is only a linear function of $\tilde{\delta}$. This approximation is only valid for infinitesimal perturbations
and severely underestimates the heat production rate as the perturbation grows towards finite amplitudes.
Nevertheless, we may assume that if the magnitude of $\tilde{\delta}$ becomes substantial before $\tilde{t}$
approaches the characteristic time $\tilde{t}_{max}$, the perturbation will develop into thermal runaway. To proceed, we therefore
investigate the growth of the quantity
\beq
\left( \frac{f_0 \left( \tilde{x}, \tilde{t}_{max} \right) }{f_0 \left( \tilde{x}, 0 \right) } \right)^{n+1}
= e^{ \left( \frac{\sigma_0}{\sigma_c} \right)^{-2} \left[ \left( \frac{\sigma_0}{\sigma_c} \right)^2
- \frac{\tau_r}{\tau_d}\pi^2 \right]^2 } \, .
\label{eq:amplify}
\eeq
We now anticipate that a thermal runaway will develop if the quantity above
amplifies beyond the characteristic factor $e$, leading to the equation
\beq
\left( \frac{\sigma_0}{\sigma_c} \right)^{-2} \left[ \left( \frac{\sigma_0}{\sigma_c} \right)^2
- \frac{\tau_r}{\tau_d}\pi^2 \right]^2  = 1 \, ,
\label{eq:maxpert}
\eeq
for which the solution is
\beq
\frac{\sigma_0}{\sigma_c} = \sqrt{\pi^2\frac{\tau_r}{\tau_d}
+ \frac{1}{2} \left[ 1 + \sqrt{1+4\pi^2\frac{\tau_r}{\tau_d}} \right] }\, .
\label{eq:thrcond}
\eeq
The expression in~(\ref{eq:thrcond}) determines the critical values of the dimensionless variables
$\sigma_0/\sigma_c$ and $\tau_r/\tau_d$ at which the transition between stable modes and unstable thermal runaway modes
occurs. Thus, at the transition between the two modes, $\sigma_0/\sigma_c$ becomes a function of $\tau_r/\tau_d$.
The curve along which the transition occurs divides the
$\sigma_0/\sigma_c$, $\tau_r/\tau_d$ plane into a ``phase diagram'', as illustrated in Fig.~\ref{fig:anstabsol}.
As the solid curve in Fig.~\ref{fig:anstabsol} shows, the critical values of $\sigma_0/\sigma_c$ increase with increasing
$\tau_r/\tau_d$. However, for $\tau_r/\tau_d < 10^{-2}$ the transition curve stays almost vertical, and the critical
$\sigma_0/\sigma_c$ approximately equals its lower bound. The lower bound is obtained in the limit
$\tau_r/\tau_d\rightarrow 0$ and corresponds to adiabatic processes. In this case Eq.~(\ref{eq:thrcond}) reduces to
\beq
\sigma_0=\sigma_c\, ,
\label{eq:adiabaticcond}
\eeq
and we thereby recover the result obtained earlier in Sec.~\ref{subsec:adtrconditions}.

Hence, $\sigma_c$ is the smallest stress required for spontaneous thermal runaway to occur. We note, however, that
the required stress for onset of thermal runaway begins to deviate significantly from $\sigma_c$
only when $\tau_r/\tau_d > 1$. Therefore $\sigma_c$ provides a good estimate for the critical stress above which
thermal runaway occurs as long as $\tau_r/\tau_d < 1$. In this regime, then, onset of thermal runaway is to a good
approximation independent of the quantities which control kinetic processes since these quantities do not appear
in the expression for $\sigma_c$ (see Eq.~(\ref{eq:criticalstress})). The results of this linear stability analysis
are consistent with earlier numerical estimates of initial stages of runaway instability in a two-dimensional
setup~\cite{Boris}. Explicit estimates of $\sigma_c$ were made in Ref.~[\onlinecite{BP}] and compared to typical
failure stresses in metallic glasses and rocks under high confining pressure, showing that the predicted values
of $\sigma_c$ are within the correct order of magnitude for such systems.

For the situation when $\tau_r/\tau_d\gg 1$ the rate of relaxation or creep in the material is very slow compared to the
process of thermal diffusion, which means that the heat produced rapidly flows away from the initially perturbed zone.
The stress required for thermal runaway to occur in this case is therefore much larger than $\sigma_c$.

Finally, it should be mentioned that the linear analysis presented in this section, and leading to
Eq.~(\ref{eq:thrcond}), was based on the greatly simplified equations (\ref{eq:linstressthree}) and
(\ref{eq:lindeltatwo}). The correctness of the obtained results must thus be investigated by a proper analysis
of the set of the fully coupled equations (\ref{eq:finstresseq}) and (\ref{eq:fintempeqtwo}). Fortunately, as will be
demonstrated in Sec.~\ref{sec:numapp}, the results of the linear analysis are found to agree extremely well
with the results obtained from numerical solutions to the complete equations.

\section{Numerical approach\label{sec:numapp}}

In order to obtain solutions to the complete equations (\ref{eq:finstresseq}) and (\ref{eq:fintempeqtwo})
without simplifying assumptions we perform numerical simulations. This enables us to capture any nonlinear effects in
our thermo-mechanical system and to study the complete time evolution of $T(x,t)$, $\sigma(t)$ and the displacement $u(x,t)$.

\subsection{Dimensionless equations\label{subsec:dimless}}

The complexity of the problem can be significantly reduced by introducing dimensionless variables. Inspired by
the analysis in sections~\ref{sec:anan} and \ref{sec:trconditions}, we choose the particular normalization
\beq
\tilde{\sigma} = \frac{\sigma}{\sigma_0}\,,\ \ \tilde{T} = \frac{T}{E/R}\,,\ \ \tilde{t} = \frac{t}{\tau_r}\,,\ \
\tilde{x} = \frac{x}{h}\,.
\label{eq:fulleqscaling}
\eeq
In most situations of interest $T_{bg}\ll E/R$ ($E/R$ is typically of the order of 10000 K), and for simplicity we
therefore restrict our numerical analysis to the case $\tilde{T}_{bg}\ll 1$.

The closed set of equations for $T$ and $\sigma$ can then be written on dimensionless form as
\beq
\frac{\partial\tilde{T}}{\partial\tilde{t}} = \frac{\tau_r}{\tau_d}\frac{\partial^2\tilde{T}}{\partial\tilde{x}^2}
+\tilde{T}_0^2 \left( \frac{\sigma_0}{\sigma_c} \right)^2 \tilde{\sigma}^{n+1}e^{\frac{1}{\tilde{T}_0}-\frac{1}{\tilde{T}}}\,,
\label{eq:nondimfulltempeq}
\eeq
and
\beq
\frac{\partial\tilde{\sigma}}{\partial\tilde{t}}
= -\frac{1}{2}\frac{h/L}{h/L+\left(1-\frac{h}{L}\right)e^{\frac{1}{\tilde{T}_0}-\frac{1}{\tilde{T}_{bg}}}}
\int_{-\frac{L}{2h}}^{\frac{L}{2h}}e^{\frac{1}{\tilde{T}_0}-\frac{1}{\tilde{T}}}d\tilde{x}\, .
\label{eq:nondimfullstresseq}
\eeq
Similarly, the dimensionless form of the rheology equation becomes
\beq
\frac{\partial\tilde{v}}{\partial\tilde{x}} = e^{\frac{1}{\tilde{T_0}}-\frac{1}{\tilde{T}}}\tilde{\sigma}^n
+ 2\Delta_p\frac{\partial\tilde{\sigma}}{\partial\tilde{t}}\,,
\label{eq:nondimrheologyeq}
\eeq
where $\tilde{v}=v/(\sigma_0h/\mu_0)$. Displacements are correspondingly calculated in units of
$(\sigma_0h/\mu_0)\tau_r=\sigma_0h/(2\Delta_pG)$.

The utility of this non-dimensionalization procedure lies in that the original problem containing thirteen dimensional
parameters now has been reduced to one containing only eight dimensionless parameters (as can be verified upon inspection
of the above equations), substantially reducing the number of necessary numerical runs. Moreover, the dimensionless
temperature $\tilde{T}$ may now be explicitly expressed as a function of the two combinations of parameters $\tau_r/\tau_d$ and
$\sigma_0/\sigma_c$ that were suggested by the linear analysis in Sec.~\ref{sec:trconditions} as controlling parameters
for onset of thermal runaway:
\beq
\tilde{T} = f_1 \left( \tilde{x},\tilde{t},\tilde{T}_0,\tilde{T}_{bg},\frac{h}{L},n,\frac{\tau_r}{\tau_d},\frac{\sigma_0}{\sigma_c}
\right) \,.
\label{eq:nondimformsol}
\eeq

To ensure correct numerical results the coupled equations (\ref{eq:nondimfulltempeq}) and (\ref{eq:nondimfullstresseq}) are
solved using a finite-difference method with non-uniform mesh and a tailored variable time-step.

\subsection{Temperature rise\label{subsec:temprise}}

Based on the analysis conducted in Sec.~\ref{subsec:dimless}, we are now in a position to investigate the thermal runaway
phenomenon in a self-consistent manner by numerical calculations of the complete time evolution of $T$ and $\sigma$
with account of heat conduction. Guided by the results obtained in Sec.~\ref{sec:anan}, we shall begin by studying the
maximum temperature rise $\Delta T_{max}$ (Eq.~(\ref{eq:defmaxtemp})), normalized by the adiabatic
temperature rise $\Delta T_{max}^a$ (Eq.~(\ref{eq:maxadtemprise})), as a function of the governing variables.

The temperature always attains the maximum value at the centre of the slab. Hence,
the maximum temperature $T_{max}$ with respect to both time and position must satisfy the equation
\beq
\left.\frac{\partial T}{\partial t}\right|_{x=0} = 0\,,
\label{eq:maxtempcond}
\eeq
from which, by use of Eq.~(\ref{eq:nondimformsol}), it is easy to show that
$\Delta\tilde{T}_{max} = \tilde{T}_{max}-\tilde{T}_0$ is independent of $\tilde{x}$ and $\tilde{t}$.
Then, by rewriting $\Delta T_{max}^a$ in terms of dimensionless variables, it is immediately clear that the
maximum temperature rise scaled by the adiabatic temperature rise is a function of the remaining six dimensionless
variables only, i.e.,
\beq
\frac{\Delta T_{max}}{\Delta T_{max}^a}
= f_2 \left( \tilde{T}_0,\tilde{T}_{bg},\frac{h}{L},n,\frac{\tau_r}{\tau_d},\frac{\sigma_0}{\sigma_c} \right) \,,
\label{eq:tmaxtmaxadratio}
\eeq
simplifying the problem even further.

\begin{figure*}[tbp]
\begin{center}
\includegraphics[width=16cm]{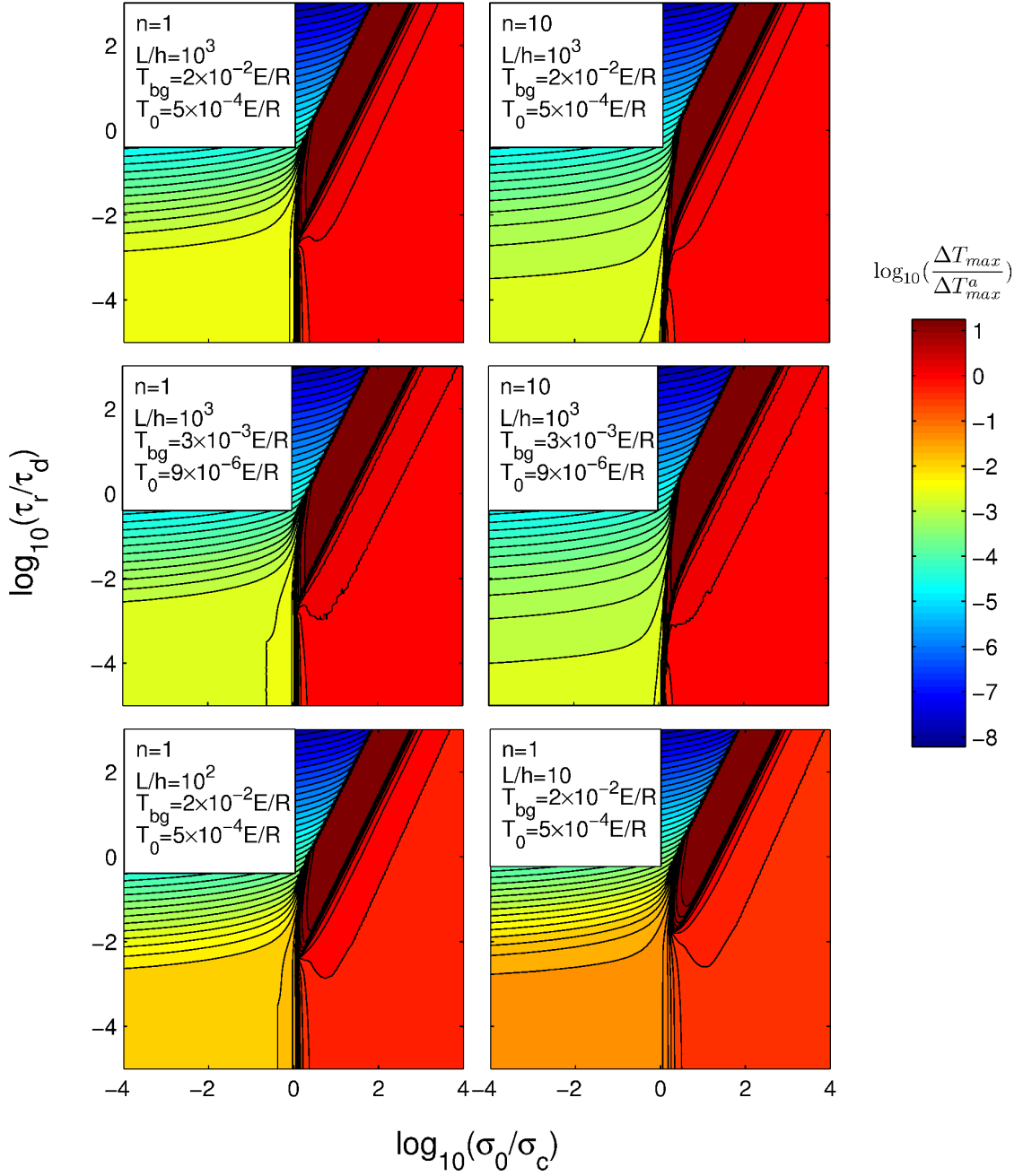}
\caption{(Color) Contour plots of $\Delta T_{max}/\Delta T_{max}^a$
versus $\sigma_0/\sigma_c$ and $\tau_r/\tau_d$. Each contour plot
represents a set of numerical runs for a certain choice of fixed
values for the parameters $n$, $L/h$, $\tilde{T}_{bg}$ and
$\tilde{T}_0$ (values are specified in white boxes). The collection
of the six contour plots thus constitutes a numerical example of
independent variations of all the six dimensionless parameters
entering Eq.~(\ref{eq:tmaxtmaxadratio}). A detailed explanation of a
contour plot is given in the caption of
Fig.~\ref{fig:Phasediagram}.} \label{fig:collapse}
\end{center}
\end{figure*}

To examine the behavior of $\Delta T_{max}/\Delta T_{max}^a$, we systematically varied all six dimensionless parameters
and computed $\Delta T_{max}$ from Eqs.~(\ref{eq:nondimfulltempeq}) and (\ref{eq:nondimfullstresseq}) for each selection
of fixed parameter values. We present several sets of numerical runs in Fig.~\ref{fig:collapse} in terms of
contour plots of $\Delta T_{max}/\Delta T_{max}^a$ versus $\sigma_0/\sigma_c$ and $\tau_r/\tau_d$ for different
values of the remaining parameters $n$, $L/h$, $\tilde{T}_{bg}$ and $\tilde{T}_0$.
As the series of contour plots show, $\Delta T_{max}/\Delta T_{max}^a$ depends strongly on the two variables
$\sigma_0/\sigma_c$ and $\tau_r/\tau_d$, but is rather insensitive to variations in the four remaining parameters.
In this sense, $\Delta T_{max}/\Delta T_{max}^a$, when plotted against the two combinations of parameters
$\sigma_0/\sigma_c$ and $\tau_r/\tau_d$, closely resembles a data collapse. It has thus been demonstrated that the
maximum temperature rise normalized by the adiabatic temperature rise is, to a quite good approximation, a function
of the two dimensionless variables $\sigma_0/\sigma_c$ and $\tau_r/\tau_d$ alone, as previously suggested by the
linear analysis (see Sec.~\ref{sec:trconditions}).

\begin{figure}[tbp]
\begin{center}
\includegraphics[width=8cm]{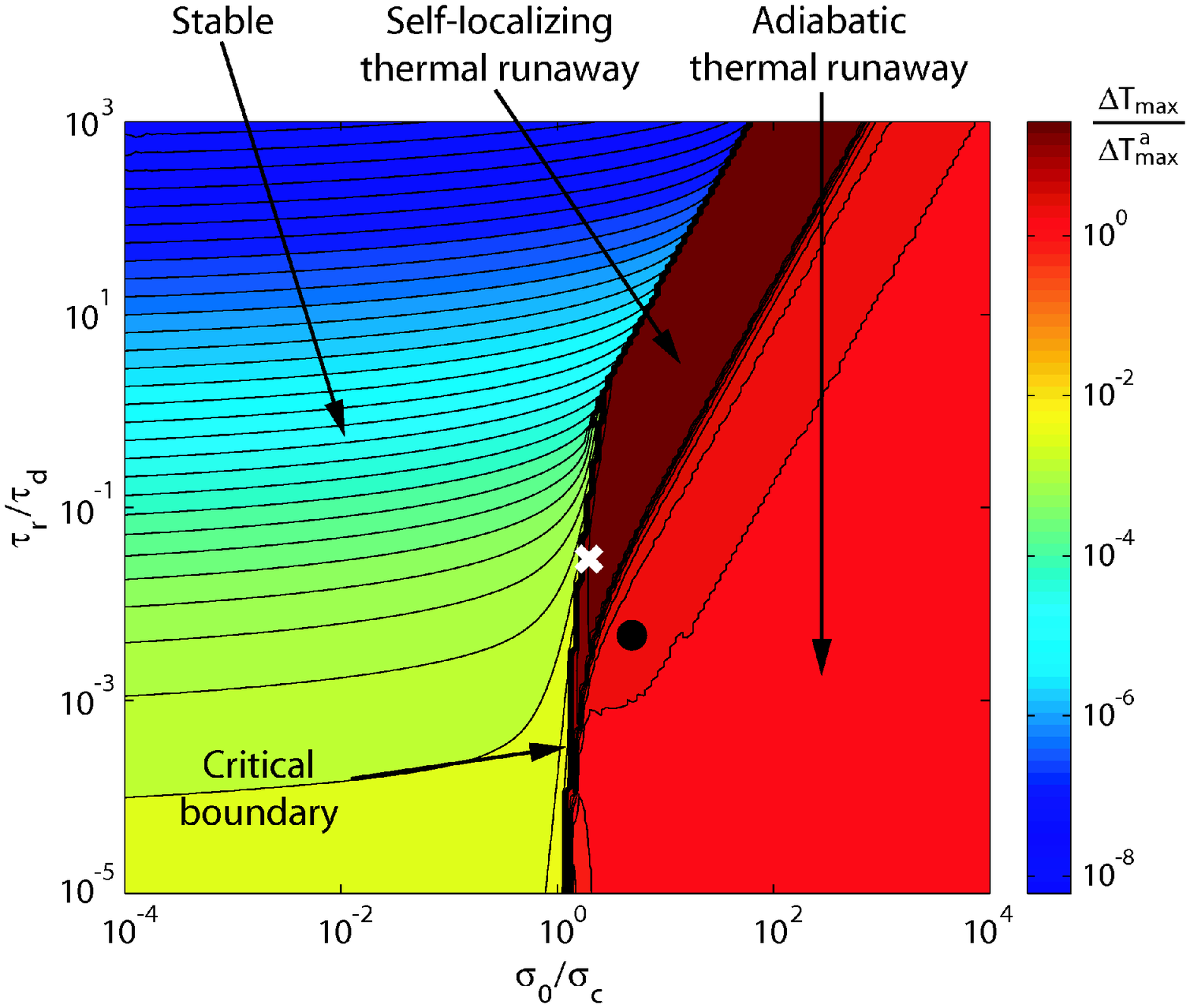}
\caption{(Color) A representative contour plot of $\Delta T_{max}$
scaled by the adiabatic temperature rise $\Delta T_{max}^a$ as a
function of the dimensionless variables $\sigma_0/\sigma_c$ and
$\tau_r/\tau_d$. The dark lines represent contour lines. The plot
exhibits mainly two sharply distinguished regions according to small
and large temperature rises, thus defining a critical boundary
dividing the $\sigma_0/\sigma_c$, $\tau_r/\tau_d$ plane into a
``phase diagram''. The low-temperature region, denoted ``stable''
(blue and green colors), represents stable deformation processes.
The high-temperature region represents unstable thermal runaway
processes and is further subdivided into two domains, denoted by
``self-localizing thermal runaway''and ``adiabatic thermal runaway''
(brown and red color, respectively), which represents runaway
processes of different characters (see text). For computational
efficiency the very late stages of the self-localizing thermal
runaway processes have not been fully resolved. The maximum
temperature rises presented for these processes are therefore
underestimated in this plot. See the captions in
Figs.~\ref{fig:adplots} and~\ref{fig:locplots} for an explanation of
the cross and the dot.} \label{fig:Phasediagram}
\end{center}
\end{figure}

Having identified the controlling variables for the maximum temperature rise in our system, we can now proceed to study
a representative contour plot, shown in Fig.~\ref{fig:Phasediagram}, in more detail.
As can be seen, the plot exhibits a low-temperature region, corresponding to stable deformation processes, and a
high-temperature region, corresponding to thermal runaway processes. As was predicted by the linear analysis in
Sec.~\ref{sec:trconditions}, these two regions are sharply distinguished by a critical boundary (or ``transition curve'')
dividing the $\sigma_0/\sigma_c$, $\tau_r/\tau_d$ plane into a ``phase diagram''. The location of the critical boundary correlates
well with the analytical predictions, see Fig.~\ref{fig:anstabsol} for comparison. This verifies that the conditions for
spontaneous thermal runaway to occur have been accurately determined. A physical explanation of the stability of the deformation
processes in the low-temperature region is that, there, the effects of thermal diffusion and stress relaxation dominate over the
positive feedback mechanism. In the high-temperature region the situation is exactly the opposite, leading to instability and
consequently thermal runaway. In the following section we turn our attention to the high-temperature region, and we shall in particular
discuss the effects of thermal diffusion on thermal runaway processes.

\subsection{Implications of thermal diffusion: Localization of deformation and temperature rise during thermal runaway
\label{subsec:localization}}

The high-temperature region exhibited in Fig.~\ref{fig:Phasediagram} is divided into two domains, as illustrated by red and
brown colors. An essential observation may be emphasized here: The maximum temperature rises in the red domain equal the adiabatic
maximum temperature rises $\Delta T_{max}^a$. However, the maximum temperature rises in the brown-colored region, i.e., in the neighborhood
of the critical boundary in the high-temperature region, are seen to be much larger than $\Delta T_{max}^a$. Therefore, these two domains
manifest thermal runaway processes of distinctly different characters, as will be outlined below.

\begin{figure}[!t]
\begin{center}
\includegraphics[width=8cm]{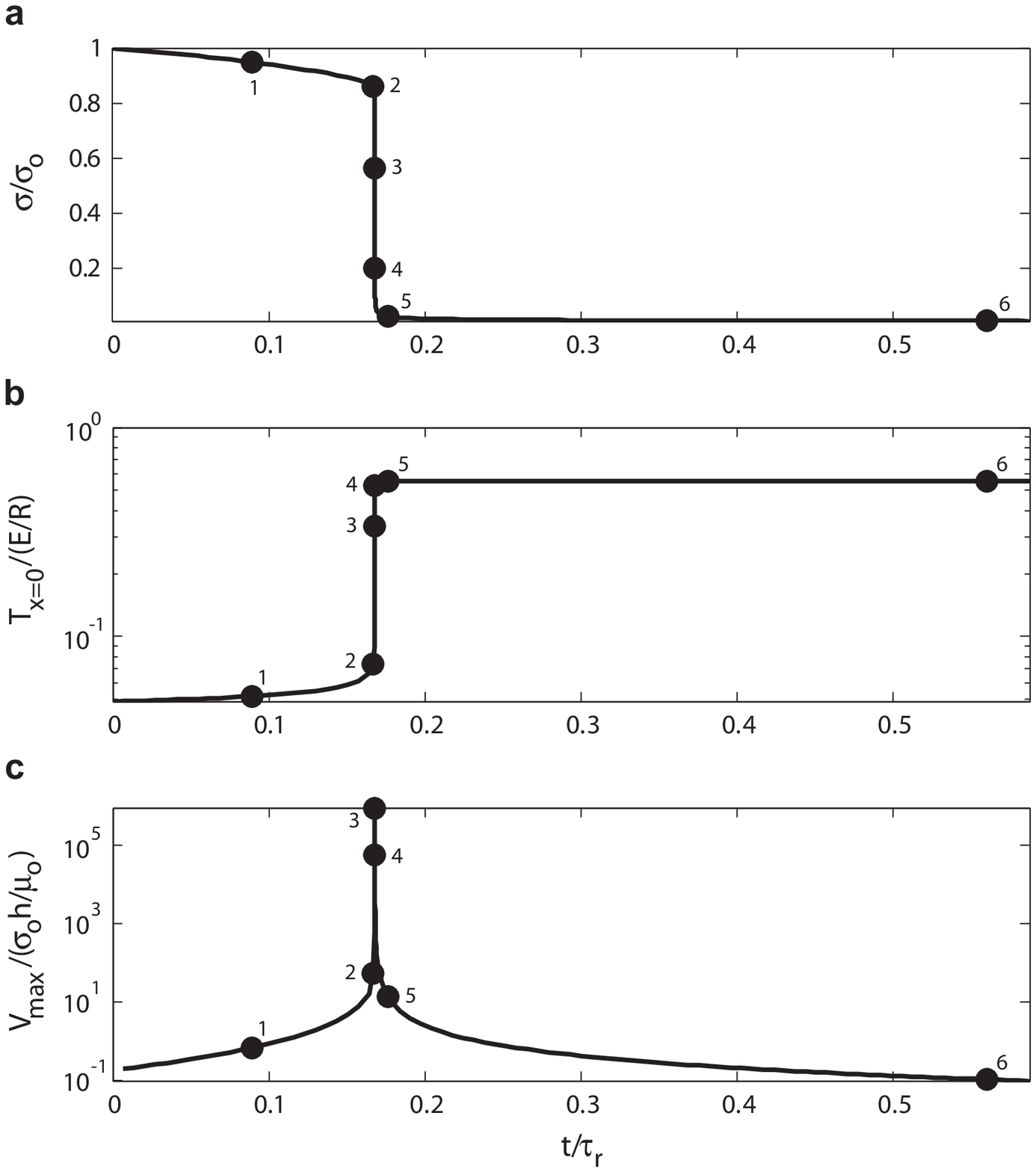}
\includegraphics[width=8cm]{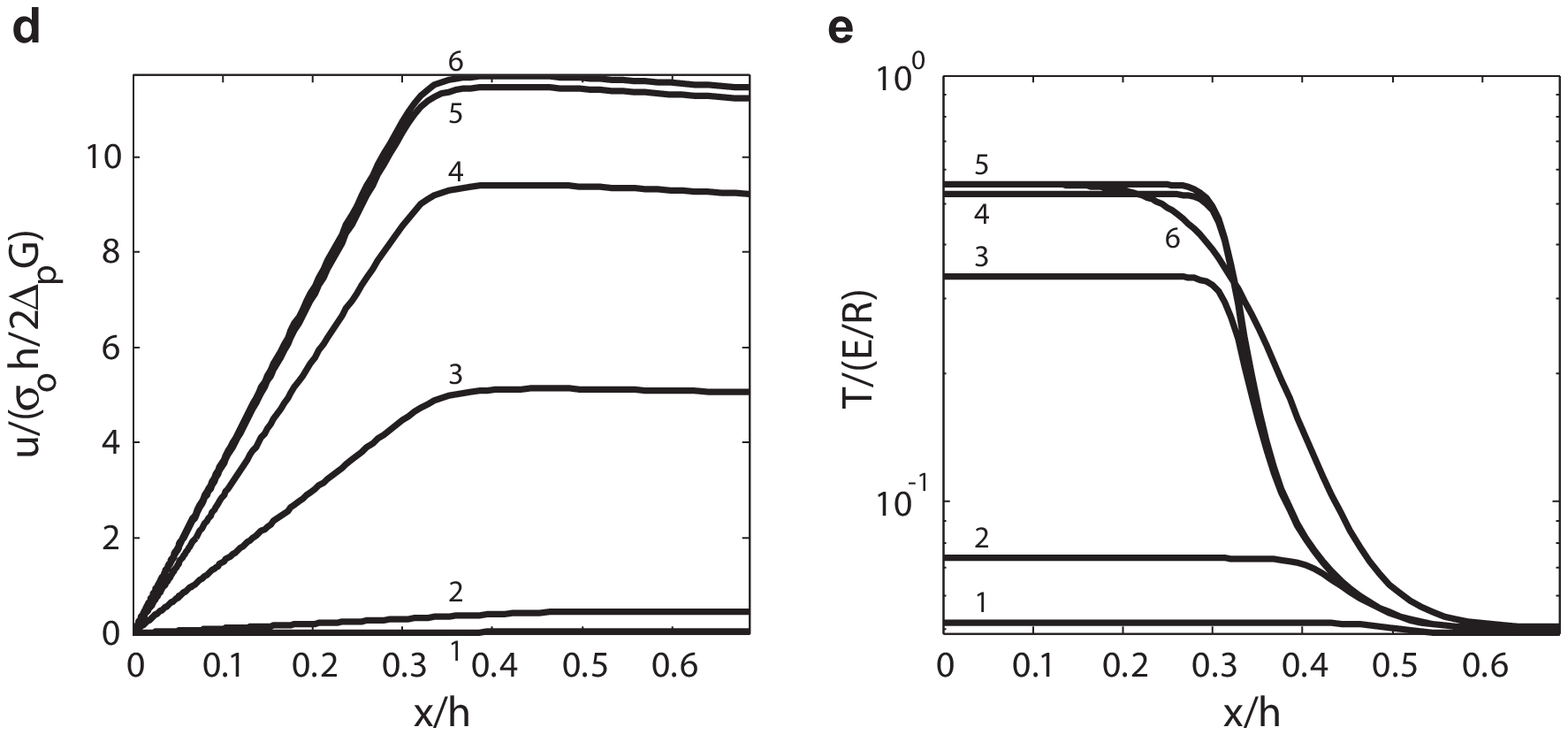}
\caption{An example of a time evolution of an adiabatic thermal
runaway process (see text) corresponding to the location of the dot
in Fig.~\ref{fig:Phasediagram}. For this particular case
$\tilde{T_0}\approx 0.0488$, $\tilde{T}_{bg}\approx 0.0476$,
$h/L\approx 0.033$, $n=4$, $\tau_r/\tau_d\approx 0.027$ and
$\sigma_0/\sigma_c\approx 1.83$. Panels (a), (b) and (c) display
respectively the shear stress $\sigma$ in units of $\sigma_0$, the
temperature in the slab center $T_{x=0}$ in units of $E/R$ and the
maximum velocity $v_{max}$ in units of $\sigma_0h/\mu_0$. The two
lower panels show a time sequence of spatial distributions in the
vicinity of the initially perturbed zone of (d) the displacement $u$
in units of $\sigma_0h/2\Delta_pG$ and (e) the temperature $T$ in
units of $E/R$ (for illustrative purposes we show profiles only for
positive $x$). The spatial distributions of $T$ and $u$ are plotted
for six different times corresponding to the time markers (black
dots) in panels (a)--(c). In all panels the time $t$ is given in
units of the relaxation time $\tau_r$, and the position $x$ is given
in units of the width $h$ of the initially perturbed zone.}
\label{fig:adplots}
\end{center}
\end{figure}

Let us first address the runaway processes occurring in the red-colored domain. An example of the time evolution of a thermal runaway
process in this domain is shown in Fig.~\ref{fig:adplots}.
As can be seen, the process may be divided into three stages
as follows. During the first stage, preceding time marker 2 (i.e., $t<0.17\tau_r$), $\sigma(t)$ is seen to decrease nearly
linearly with time while the temperature in the slab center $T_{x=0}(t)$ and the maximum velocity $v_{max}(t)$ increase gradually.
A corresponding gradual increase in the spatial distributions of the temperature $T$ and the displacement $u$ is noticed in the
outer part of the initially perturbed zone having width $h$.
However, during the second stage, covered by time markers 2--5 (i.e., in the neighborhood of $t\approx 0.17\tau_r$), the shear stress
$\sigma$ spontaneously drops to zero and is accompanied by a corresponding explosive rise in the temperature $T_{x=0}$ up to a
maximum, whereas one observes an explosive increase in the maximum velocity $v_{max}$ up to a peak value immediately followed by a rapid
decrease to a much smaller value again. Moreover, during this short period of time, a dramatic increase in the displacement $u(x,t)$ has
occurred in the outer part of the initially perturbed region, while a major, essentially uniform rise in the
temperature $T(x,t)$ is observed in the inner part of this region ($0\leq x < 0.4h$). This stage, then, represents an extremely rapid
thermal runaway process occurring during a time interval much smaller than the thermal diffusion time. The effects of thermal
diffusion are therefore seen to be negligible. Finally, during the third stage, succeeding time marker 5 (i.e., $t>0.17\tau_r$),
$v_{max}$ decreases smoothly towards zero while essentially no further changes in the remaining quantities can be seen.
It is worth emphasizing that the rather constant value of the temperature $T_{x=0}$ at this last stage stems from the thermal diffusion
time being too large for any noticeable effects of thermal diffusion to be seen within the small time interval exhibited in these plots.

In summary, we observe that thermal runaway processes in the red-colored domain are characterized by a rate of heat production
which is much greater than the rate of heat conduction. Accordingly, the runaway process is well approximated as adiabatic and, as was
explained in Sec.~\ref{subsec:adtemp}, the elastic energy in the slab is therefore dissipated essentially uniformly throughout the
initially perturbed zone ($|x|\leq h/2$), thus producing a maximum temperature rise $\Delta T_{max}$ which equals the adiabatic
temperature rise $\Delta T_{max}^a$. Moreover, during the runaway process a dramatic increase in the displacement occurs, which
represents the formation of an adiabatic shear band having width of the same order of magnitude as the width $h$ of the initially
perturbed zone.

\begin{figure}[tbp]
\begin{center}
\includegraphics[width=8cm]{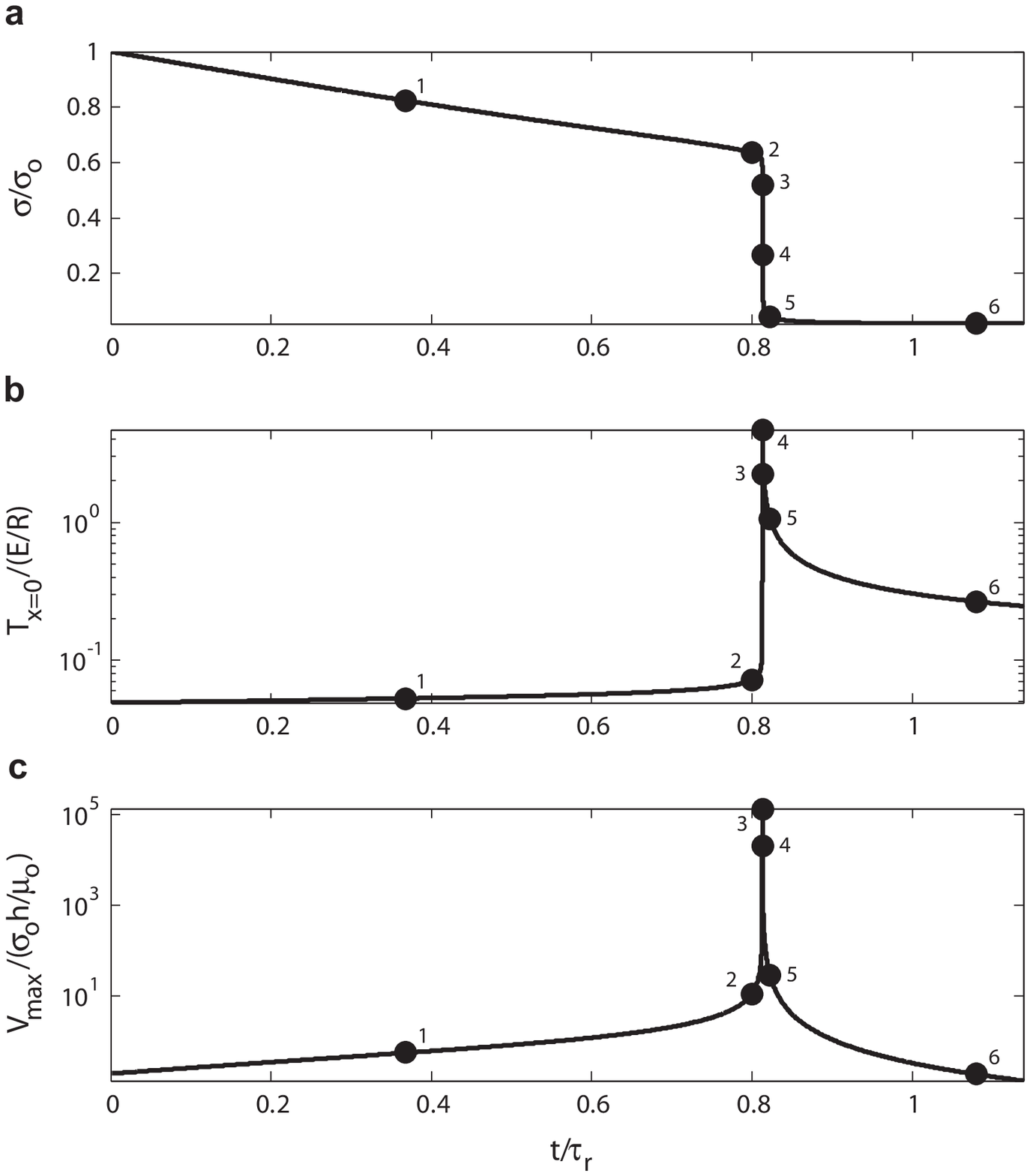}
\includegraphics[width=8cm]{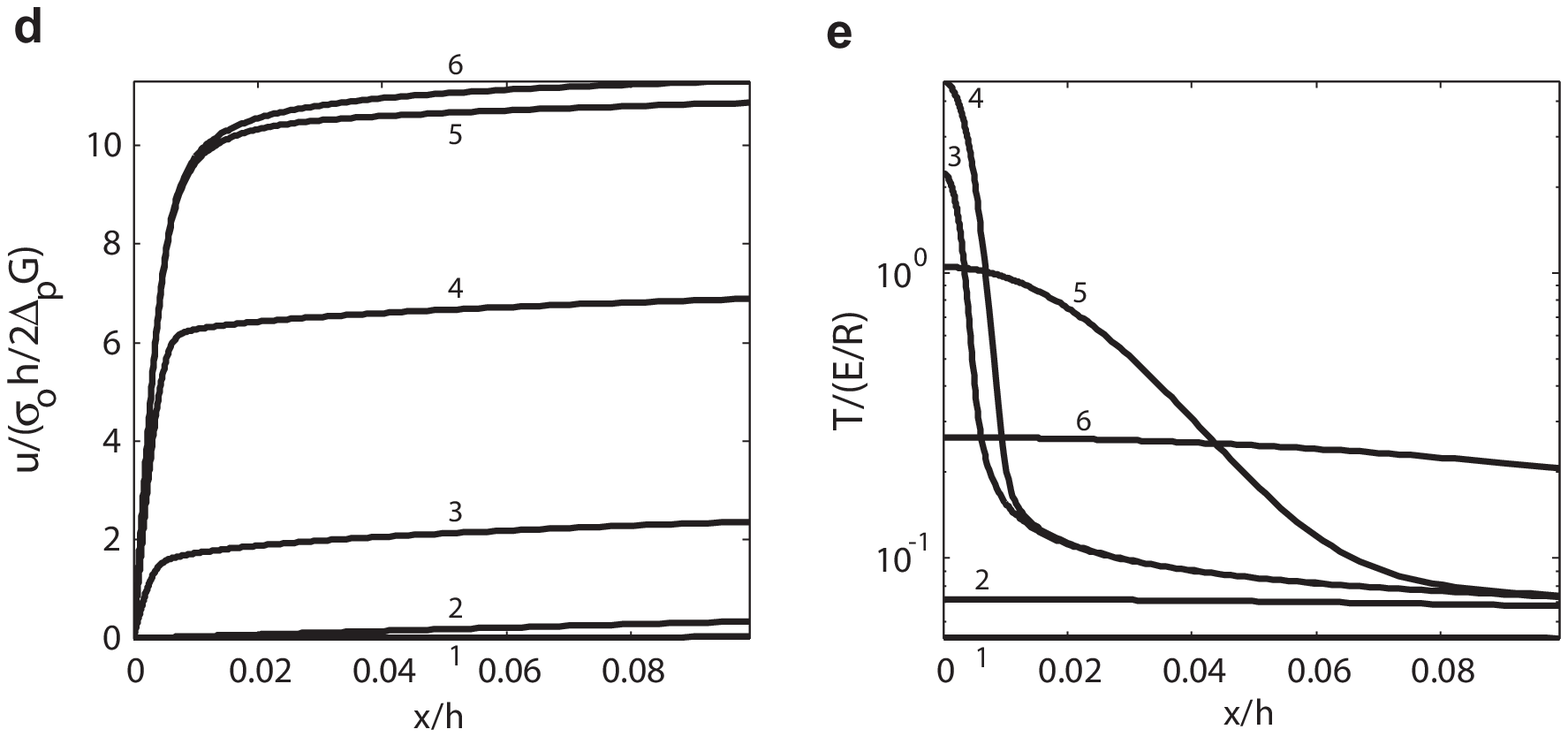}
\caption{An example of a time evolution of a self-localizing thermal
runaway (see text) corresponding to the location of the cross in
Fig~\ref{fig:Phasediagram}. Here $\tilde{T_0}\approx 0.0488$,
$\tilde{T}_{bg}\approx 0.0476$, $h/L\approx 0.033$, $n=4$,
$\tau_r/\tau_d\approx 0.006$ and $\sigma_0/\sigma_c\approx 3.06$.
Panels (a)--(e) are organized in the same manner as outlined in the
caption of Fig.~\ref{fig:adplots}. Note, however, that here the
spatial scale in panels (d) and (e) is much smaller than the spatial
scale of the corresponding panels (d) and (e) in
Fig.~\ref{fig:adplots}.} \label{fig:locplots}
\end{center}
\end{figure}

Next, we proceed to analyze the thermal runaway processes occurring in the brown-colored domain, exhibiting much larger temperature rises
than $\Delta T_{max}^a$. Figure~\ref{fig:locplots} illustrates an example of a thermal runaway process in this domain.
Once again, the time evolution of the process may be divided into three stages; stage one precedes time marker 2 (i.e., $t<0.8\tau_r$),
stage two is covered by markers 2--5 (i.e., $t\approx 0.8\tau_r$), whereas the last stage succeeds time marker 5
(i.e., $t>0.8\tau_r$). The initial stage is characterized by a relatively large time interval in
which the stress $\sigma$ decreases approximately linearly with time. During the same stage it is seen that the temperature
$T_{x=0}$ and the maximum velocity $v_{max}$ increase gradually. There is basically no change in the spatial
distributions of the displacement $u(x,t)$ or in the temperature $T(x,t)$. During the second stage, however, a thermal runaway
occurs and thus the stress declines spontaneously towards zero. Along with this spontaneous stress drop, one observes an explosive rise in
the central temperature up to a maximum followed by a very rapid decrease. It should be emphasized here that, in contrast to what was observed
in Fig.~\ref{fig:adplots}, the later decrease in $T_{x=0}$ highlights the importance of thermal diffusion for this particular type of thermal
runaway. The maximum velocity is seen to accelerate extremely fast up to a peak value immediately followed by an equally fast decrease to a much
smaller value again. The dramatic rise in $v_{max}$ as one moves from stage one to stage two is accompanied by a corresponding major rise
in $u(x,t)$ and $T(x,t)$. However, the displacement and temperature profiles in this case show strikingly different features as compared to the
same profiles obtained in the former case that was shown in Fig.~\ref{fig:adplots}. Indeed, in this case, large deformation
is seen to occur much closer to the slab center and an extremely non-uniform rise in temperature occurs around the origin. In other words,
the strain and temperature profiles continuously localize during the rapid deformation process. Finally, during the third stage,
$T_{x=0}$ decreases gradually due to thermal diffusion and the deformation process terminates as $v_{max}$ approaches zero.
\begin{figure}[tbp]
\begin{center}
\includegraphics[width=8cm]{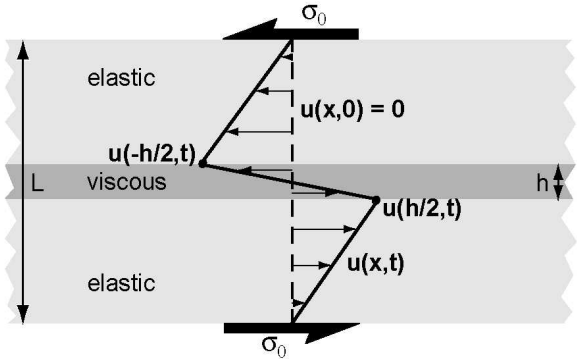}
\caption{Two rigid elastic plates sliding past one another on a thin
viscous layer.} \label{fig:slidingslabs}
\end{center}
\end{figure}

In conclusion, we have seen that thermal runaway processes in the brown-colored domain are characterized by continuous localization
of the temperature and strain profiles during deformation, i.e., these runaway processes are spatially ``self-localizing''. The elastic
energy is thus dissipated in a zone much narrower than the width of the initial perturbation, resulting in maximum temperature rises
$\Delta T_{max}$ which are much larger than the maximum adiabatic temperature rises $\Delta T_{max}^a$. The self-localization of
these runaway processes clearly arises from the effects of thermal diffusion: by diffusion the temperature profile initially acquires
a peak in the center where the effect of the positive feedback mechanism accordingly is maximized. The runaway therefore develops
faster in the center than in the regions outside and the deformation process finally terminates in a highly localized shear band
with a characteristic width much smaller than the width $h$ of the initially perturbed zone.

Finally, we emphasize that the self-localizing failure modes occur at lower values of the shear stress compared to the adiabatic modes.
Hence, if the material is subjected to a shear stress large enough to initiate a thermal runaway (i.e., $\sigma\sim\sigma_c$), the failure
process is expected to be nonadiabatic and to involve a continuous thinning of the developing shear band.

\section{Relation between evolution of stress, displacement and temperature rise\label{sec:stressdrop}}

Thus far the possibility of catastrophic material failure by spontaneous thermal runaway has been demonstrated. An interesting question
yet to be investigated, however, is the relation between stress drop, displacement, shear-band width and temperature rise during runaway,
as now discussed.

For this purpose, let us again consider the viscoelastic slab shown in Fig.~\ref{fig:modelsetup}. Assume that the viscosity in the
central region $|x|\leq h/2$ is independent of $x$ and much smaller than in the regions outside. In this case the system behaves
as if two rigid elastic plates slide past one another on a thin viscous layer (recall, however, that the deforming slab is not
far-field driven, but that the outer boundaries $x=\pm L/2$ are clamped as internal deformation occurs). This situation is illustrated in
Fig.~\ref{fig:slidingslabs} where we have defined the initial distribution of the displacement field $u(x,t)$ to be $u(x,0)=0$.
In accordance with this definition, the shape of the displacement profile is as shown in the figure, and we may define a
relative displacement $u_r(t)=u(h/2,t)-u(-h/2,t)$ between the boundaries of the viscous layer.
For simplicity, we shall study the case when the width of the viscous layer $h$ is much smaller than the system size $L$, i.e.,
$h/L\ll 1$. Then, denoting the initial stress by $\sigma_i$, it is clear that the maximum possible relative displacement is approximately
given by
\beq
u_r^{max}\approx\frac{L}{G}\sigma_0 \, .
\label{eq:maxreldisp}
\eeq
We stress here that $\sigma_0$ is the initial shear stress in the slab, whereas $u_r^{max}$ is the displacement corresponding to
the final state for which the slab is completely unloaded.
As the rigid plates continue to slide past one another elastic energy stored in the rigid plates is continuously dissipated
as heat in the viscous layer. If the sliding process occurs in a time short compared to that for thermal diffusion,
the energy equation for this layer may be written
\beq
C\frac{\partial T}{\partial t} = \sigma(t)\frac{\partial\gamma}{\partial t}\, ,
\label{eq:sdtempeq}
\eeq
where $\gamma$ denotes the strain inside the viscous central region and the term on the r.h.s. of the equation is the work dissipated
during irreversible viscous flow. It was seen in the examples of thermal runaway events in the
previous section that the shear stress does not remain constant during deformation, but drops spontaneously towards zero. Hence,
$\sigma(t)$ is a strongly varying function of time and as a consequence Eq.~(\ref{eq:sdtempeq}) cannot be integrated over time
directly. This apparent complication may be circumvented, however, by instead expressing both $\sigma$ and $\gamma$ as
functions of the displacement. In correspondence with the adiabatic assumption made here, the temperature rise within the viscous
layer will be uniform. Since the viscosity then is a function of both a uniform temperature and shear stress, it follows that
the viscosity and accordingly the strain within the viscous layer is uniform. Then, from inspection of Fig.~\ref{fig:slidingslabs},
it is seen that the uniform strain can be expressed as a function of the relative displacement $u_r$ according to the relation
\beq
\gamma(t) = \frac{u_r(t)}{h}\, ,
\label{eq:vstrain}
\eeq
from which we obtain the strain rate
\beq
\frac{d\gamma}{dt} = \frac{1}{h}\frac{du_r(t)}{dt}\, .
\label{eq:vstrainrate}
\eeq
Similarly, in the limit $L>>h$ and because the outer rigid plates are purely elastic, it can be gleaned from
Fig.~\ref{fig:slidingslabs} that the stress approximately is given by
\beq
\sigma(t) \approx\sigma_0 - G\frac{u_r(t)}{L}\, .
\label{eq:estress}
\eeq
Substituting these expressions for $d\gamma/dt$ and $\sigma(t)$ in Eq.~(\ref{eq:sdtempeq}) and using that
$u_r(t)(du_r/dt)=1/2(du_r^2/dt)$, the energy equation takes the form
\beq
C\frac{dT}{dt} = \frac{d}{dt} \left[ \left( \sigma_0-\frac{G}{2L}u_r(t) \right) \frac{u_r(t)}{h} \right] \,.
\label{eq:derivtempdisp}
\eeq
Next, it follows from Eq.~(\ref{eq:estress}) that $(G/2L)u_r(t)=\frac{1}{2}(\sigma_0-\sigma(t))$, and the energy equation
may thus be written
\beq
C\frac{dT}{dt} = \frac{d}{dt} \left[ \frac{1}{2} \left( \sigma_0+\sigma(t) \right) \frac{u_r(t)}{h} \right] \,,
\label{eq:elimrel}
\eeq
thereby eliminating $G/L$ from the equation. The desired relation between dynamic quantities can now be obtained by direct integration
of the energy equation with respect to time. If the deformation process terminates at some final time $t_f$, then upon integrating
from $t=0$ to $t_f$, we find
\beq
C\Delta T^f = \frac{(\sigma_0+\sigma_f)}{2}\frac{u_r^f}{h} \, .
\label{eq:stresstemprel}
\eeq
Here $\sigma_f=\sigma(t_f)$, $\Delta T^f=T(x,t_f)-T(x,0)$ and $u_r^f=u_r(t_f)$.
Not surprisingly, Eq.~(\ref{eq:stresstemprel}) shows that to correctly account for decrease in shear stress during deformation,
one should use the average of the initial and final shear stress in calculating the work done by the rigid plates on the viscous layer.

The relation between dynamic quantities in Eq.~(\ref{eq:stresstemprel}) was obtained under the assumption of adiabatic conditions.
Yet, it was clarified in the previous section that thermal runaway processes are greatly affected by thermal diffusion and therefore
not truly adiabatic. As a consequence shear was seen to localize to a region of width much smaller than the width $h$ of the initially
perturbed region. It is therefore of interest to evaluate the accuracy of Eq.~(\ref{eq:stresstemprel}) for the more general
case, including self-localizing runaway processes, by numerical methods. For this reason proper definitions of the quantities entering
Eq.~(\ref{eq:stresstemprel}), valid also for nonadiabatic processes, are needed.

Figure~\ref{fig:locplots} shows that the first stage of stress relaxation does not contribute notably to deformation. We define
the initial shear stress $\sigma_i$ associated with the process of shear-band formation to be the stress at the instant $t_i$ at which the
curvature of the temporal stress curve $d^2\sigma/dt^2$ first becomes negative. Thus $\sigma_i$ is the shear stress in the slab right at
the onset of thermal runaway. The final stress, $\sigma_f$, is defined as the stress corresponding to the instant $t_f$ at which
$d^2\sigma/dt^2$ exhibits a maximum. $\sigma_f$ is thus the shear stress in the slab right at the instant at which the thermal runaway process terminates.
For illustration, $\sigma_i$ and $\sigma_f$ are the stresses corresponding to time markers 2 and 5 in Fig.~\ref{fig:locplots}(a),
respectively. Next, in order to define the shear band properly, we must identify the point in the displacement profile where a sharp
transition from very large strain to small strain occurs (for example, in the displacement profile corresponding to time marker 6 in
Fig.~\ref{fig:locplots}(d), this sharp transition occurs at the position $x/h\approx 0.01$). For this reason, we must analyze the second
derivative of the displacement profile $\partial^2u/\partial x^2$ at the instant of time when the runaway process terminates
(essentially no further deformation occurs for later times). We define the shear band to be the region $|x| \leq w/2$, where $x=w/2$
corresponds to the position at which $\partial^2u/\partial x^2$ becomes a minimum (by symmetry $\partial^2u/\partial x^2$ becomes a maximum
at $x=-w/2$). The final relative displacement across the shear band is then given by $u_r^f=u(w/2,t_f)-u(-w/2,t_f)$. In our numerical
calculations we replace the quantities $\sigma_0$ and $h$ in Eq.~(\ref{eq:stresstemprel}) by $\sigma_i$ and $w$, respectively. Lastly,
the temperature rise $\Delta T^f$ in Eq.~(\ref{eq:stresstemprel}) is now taken to be the maximum temperature rise $\Delta T_{max}$ occurring
in the center of the slab.

With all the quantities properly defined, and since the controlling variables for onset of thermal runaway have been identified, it is
now possible to investigate the robustness of Eq.~(\ref{eq:stresstemprel}) by numerical analysis.
Numerical calculations of the quantities entering Eq.~(\ref{eq:stresstemprel}) versus the controlling physical
parameters $\sigma_0/\sigma_c$ and $\tau_r/\tau_d$, where the controlling parameters are varied over the same range as in
Fig.~\ref{fig:Phasediagram}, show that
\beq
1 \lesssim \frac{2wC\Delta T_{max}}{\left( \sigma_i + \sigma_f \right) u_r^f} \lesssim 2 \,.
\label{stresstemprelcheck}
\eeq
Hence, according to our model, the simple relation in~(\ref{eq:stresstemprel}) is applicable even to nonadiabatic situations
leading to self-localizing processes.

\section{Discussion and conclusions\label{sec:conc}}

The mechanism of spontaneous thermal runaway in viscoelastic solids
has been analyzed within the framework of a highly simplified
one-dimensional continuum model. As a first approximation the model
is only intended to contain the most essential physics governing the
thermal runaway phenomenon and certainly neglects many important
effects encountered in the more complicated experimental situations.

Among the most crucial simplifications made in the present model is
perhaps our neglect of significant changes of material properties
that may result from the very large temperature rises associated
with the runaway instability. The model does not include corrections
due to effects of melting, although melting of the material near the
shear band may be possible. Also, the influence of melting on
shear-band formation could potentially increase the rate of
deformation even further and cause propagation of elastic waves,
which would then require consideration of inertial effects. For
these reasons the calculations made within the present model are
strictly valid only until these changes take place, and the model
may therefore be assumed to reproduce only qualitatively the correct
deformation behavior during the later stages of the thermal runaway
process. However, the inertia becomes significant only after the
onset of localization and/or thermal runaway and hence does not
affect the conclusions of our study. This statement was also checked
by additional numerical studies based on the full system of
equations which include the inertial terms.

Our approach has thus been to investigate the simplest possible
model which still has some expectation of representing a real
physical situation reasonably well. Although the ideal character of
the model should not be ignored, it allows for a quantitative
treatment of the deformation problem that hopefully provides
valuable information about the behavior of the thermal runaway
failure mechanism.

A basis for the theory is the assumption that the temperature dependence of the material's viscosity can be described approximately
by an Arrhenius expression, and that it generally has a nonlinear dependence on the shear stress. This represents a simple model
accounting for thermally activated transitions in the solid, believed to be responsible for nonelastic behavior below the yield
stress. As a consequence, thermal runaway instability may occur due to shear heating-induced thermal softening of the material.

In order to determine the conditions necessary for thermal runaway to occur a theoretical analysis was carried out. Neglecting the effect
of thermal diffusion, i.e., assuming adiabatic conditions during deformation, approximate analytical solutions supposed to be valid for
the early stages of time evolutions of temperature and shear stress were found. For this case it was shown that the system becomes unstable
against shear banding due to spontaneous thermal runaway if the initial shear stress $\sigma_0$ becomes larger than a critical stress
$\sigma_c$ (Eq.~(\ref{eq:criticalstress})). To investigate the effects of thermal diffusion on the instability conditions, we subsequently
performed a linear analysis using rather rough approximations of the complete system of equations. The stability condition obtained within
the adiabatic approximation was then modified to include an additional controlling combination of physical parameters, namely the ratio of
the relaxation time $\tau_r$ to the thermal diffusion time $\tau_d$. According to this analysis $\sigma_c$ still provides a good estimate
for the critical stress above which spontaneous thermal runaway occurs as long as $\tau_r/\tau_d < 1$. However, when $\tau_r/\tau_d\gg 1$,
the rate of relaxation or creep in the material is very small compared to the process of thermal diffusion, and the stress required for
thermal runaway to occur in this case is therefore much larger. These results were independently verified by finite
amplitude numerical analysis as demonstrated by a series of contour plots of the maximum temperature rise
$\Delta T_{max}$~(Fig.~\ref{fig:collapse}). Hence we conclude that initiation of spontaneous thermal runaway is controlled by the two
combinations of parameters $\sigma_0/\sigma_c$ and $\tau_r/\tau_d$ only.

Numerical investigation of the maximum temperature rises and displacement profiles during thermal runaway instabilities revealed
two potential types of thermal runaway processes having distinctly different characters. The first, occurring under near adiabatic
conditions, is characterized by essentially uniform temperature rise and strain inside the perturbed zone and is therefore referred to as
adiabatic thermal runaway. The second, occurring under nonadiabatic conditions, is characterized by continuous localization of the
temperature and strain profiles during deformation and is accordingly referred to as self-localizing thermal runaway.
However, the self-localizing failure modes occur at lower values of the shear stress compared to the adiabatic modes. In materials
subjected to increasing loads the actual failure process is therefore expected to be nonadiabatic. Thus, if the shear stress in the material
exceeds a critical value of the order of $\sigma_c$, the material starts to internally disintegrate by unloading the elastic energy stored
in the bulk of the medium through accelerated creep along a continuously narrowing band. Since creep is a thermally activated process, this
rapid increase in creep is achieved by local rise in temperature. In turn, the accelerated creep prevents the material from local cooling due
to thermal conduction. This opens for the possibility that some materials become unstable against macroscopic perturbations (that localize
extremely while developing) before reaching the theoretical shear strength limit at which the material would break locally, i.e., at the
lattice scale. In this way the material may fail by ductile deformation at scales much smaller than the deforming sample size, and notably at
scales much smaller than the characteristic width of an initial thermal perturbation, but which still are orders of magnitude larger than
interatomic spacing.

Finally, a quantitative relation between evolution of stress, deformation and temperature rise was obtained for adiabatic shear
banding processes by analytical methods. Numerical calculations of self-localizing thermal runaway processes within our simple
viscoelastic model have been carried out, showing that the same relation is valid also for this particular case. In order to
establish the robustness of Eq.~(\ref{eq:stresstemprel}), however, it would be instructive to compare this relation to the results
of improved model calculations of the later stages of thermal runaway processes taking into account important changes in material
properties as commented upon above.

Recent studies~\cite{Torgeir08,pseudothac} showed that the theory
developed here and in Ref.~[\onlinecite{BP}] is applicable to
explain the generation of intermediate-depth earthquakes. In
contrast to this study in which we consider infinitesimal
perturbations ($\Delta_p\approx1$ and $T_{0} \approx T_{bg}$), John
et al.~\cite{pseudothac} considers large amplitude finite
perturbations of the system caused by water influence on rheological
properties of rocks subjected to differential stresses. Even though
the finite amplitude perturbations distort the data collapse (such
as presented in Fig.~\ref{fig:collapse}), the representation of
results as function of the two combinations of parameters
$\sigma_0/\sigma_c$ and $\tau_r/\tau_d$ was proved to be useful.
Using laboratory derived properties of diabase, the typical
representative of lower crust rocks, John et
al.~[\onlinecite{pseudothac}] show that the self-localizing thermal
runaway can be considered as a potential mechanism for deep
earthquakes.

A thorough, quantitative comparison of the theory presented here with
experiments on bulk metallic glasses and polymers at various loading conditions and temperatures is
outside the scope of the present discussion. Nevertheless, a very rough estimate of the ratios
$\sigma_0/\sigma_c$ and $\tau_r/\tau_d$ for bulk metallic glasses is possible.
For bulk metallic glasses, typical values are
$C$=1.6$\times10^6$ J~m$^{-3}$K$^{-1}$
(Ref.~[\onlinecite{Lewandowski}]), $G$=34 GPa (Ref.~
[\onlinecite{JohnsonSamwer}]), $E$= 100--400 kJ~mole$^{-1}$
(Ref.~[\onlinecite {Wang,Chen}]). Assuming a temperature $T_{bg}\approx
620$ K and infinitely-small amplitude of the perturbation, i.e., $\Delta_p\approx 1$, we obtain
$\sigma_c$ = 0.8--2 GPa. Our study shows, however, that the critical value of the stress
$\sigma_0$ needed to initiate self-localizing thermal runaway
may differ significantly from $\sigma_c$ if the ratio
$\tau_r/\tau_d$ is large. The estimate of this ratio is much
less certain. Only the thermal diffusivity $\kappa$ has a
well-established value of 3$\times 10^{-6}$ m$^2$s$^{-1}$
(Ref.~[\onlinecite{Lewandowski}]). From Ref.~[\onlinecite{Johnson}] the viscosity $\mu_0$ at this
temperature is inferred to be approximately 10$^{11}$--10$^{12}$ Pa$\cdot$s.
The characteristic width $h$ of the initial perturbation, representing a
macroscopic perturbation in the material, is highly uncertain but a lower
bound may be estimated. Experimental observations~\cite{Lewandowski} of the
width $w$ of the zone affected by deformation around shear bands give
$w\approx 1$ $\mu$m. Because of the self-localizing nature of the deformation
(Fig.~\ref{fig:adplots}c), we expect that $h\gg w$, and hence
we assume $h$ = 10 $\mu$m - 10 mm (limited by the typical sample size).
These very rough estimates give a value for the dimensionless ratio $\tau_r/\tau_d$
ranging from 10$^{-2}$ to 10$^5$ and the critical value of $\sigma_0$ necessary for
self-localizing thermal runaway to occur may be expected to be close to
the critical stress $\sigma_c$. It should be noted that our estimate of
$\sigma_0\approx\sigma_c\approx$ 0.8-2 GPa represents the upper limit of critical stress
in the system. Modifications of our idealized model setup by e.g. increasing the intensity
of the initial perturbation (decreasing $\Delta_p$) and accounting for the effect of
dynamically building up the shear stress in the system would significantly decrease the
stress required to initiate instability~\cite{pseudothac}.
Without detailed analysis of particular cases, a quantitative comparison of our theory
to experimental results on bulk metallic glasses is, at present, somewhat difficult as the above
estimate of the ratio $\tau_r/\tau_d$ clearly is insufficiently constrained. Explicit studies
invoking appropriate constitutive behavior should be undertaken; for example, it might be more
realistic using a Vogel-Fulcher-Tamman or Cohen-Grest~\cite{CohenGrest} dependence of viscosity
on temperature instead of the Arrhenius dependence employed here. However, the self-localizing
thermal runaway mechanism appears to be compatible with current experimental data and should,
in our opinion, be considered as a potential mechanism governing instabilities also in materials
such as bulk metallic glasses and glassy polymers.

The study~\cite{pseudothac} also demonstrated that the theory may be
applicable to the system described not only for thermal, but also
for rheological perturbations such as variations of activation
energy of creep $E$ and/or pre-exponential constant $A$. The
rheological model of the system (Eq.~\ref{eq:viscosity}) may also be
extended to include low temperature plasticity (Peierl´s
plasticity~\cite {Schulson}).

\acknowledgments{This work was supported in part by the Norwegian
Research Council through a Center of Excellence grant to PGP.}

\end{document}